\documentclass{aa}  

\usepackage{graphicx}
\usepackage{float} 
\usepackage{txfonts}
\usepackage{lipsum}
\usepackage{subcaption}         
\usepackage{lscape}             
\usepackage{placeins}            
\usepackage{natbib}
\bibpunct{(}{)}{;}{a}{}{,}      
\usepackage[colorlinks=true, linkcolor=blue, citecolor=blue, urlcolor=blue]{hyperref}

\usepackage{caption}
\begin{document}
\authorrunning{L. Coulaud et al.}
   \title{Experimental investigation of O\textsubscript{2} diffusion and entrapment in interstellar amorphous solid water (ASW)}

   \subtitle{}


   \author{L. Coulaud\inst{1,2}\fnmsep\thanks{Corresponding author: lina.coulaud@orange.fr}, J. C. Santos\inst{2,3}\thanks{51 Pegasi b Fellow}, K.-J. Chuang\inst{2}
        }

   \institute{Département de Chimie, Université Paris-Saclay, Gif-sur- Yvette, 91190, France
    \and Laboratory for Astrophysics, Leiden Observatory, Leiden University, PO Box 9513, 2300 RA Leiden, The Netherlands
    \and Center for Astrophysics, Harvard \& Smithsonian, 60 Garden St., Cambridge, MA 02138, USA\\ }

   \date{Received date / Accepted date}

 
  \abstract
   {Interstellar ices are mainly composed of amorphous solid water (ASW) containing small amounts of hypervolatiles, such as O\textsubscript{2}, whose diffusion-limited reactions play a key role in space chemistry. Although O\textsubscript{2} is an important precursor molecule present during the early stages of ice formation, its surface diffusion in ASW remains poorly constrained.}
   {In this study, we experimentally investigate the surface diffusion and the entrapment efficiency of O\textsubscript{2} in porous ASW under astrophysically relevant conditions.} 
   {Experiments were conducted in an ultra-high vacuum chamber and monitored using infrared (IR) spectroscopy and quadrupole mass spectrometry. Diffusion coefficients were extracted through a novel approach applicable to IR-inactive molecules, by fitting the mass spectrometer signal during the isothermal phase with a Fickian model. These coefficients were then used to derive the diffusion energy barrier of O\textsubscript{2} in ASW. Entrapment efficiencies were measured by analyzing the subsequent temperature-programmed desorption phase.}
   {We measured the surface diffusion coefficients at different temperatures (35 K, 40 K, 45 K) and water ice coverages (40 ML, 60 ML, 80 ML), yielding values on the order of 10\textsuperscript{$-$16}$-$10\textsuperscript{$-$15} cm\textsuperscript{2} s\textsuperscript{$-$1}. From these values, we derived a diffusion energy barrier of $E_{\mathrm{D}}$ = 10 $\pm$ 3 meV (116 $\pm$ 35 K), corresponding to a $\chi$ ratio of about 0.1. Entrapment measurements revealed that a residual amount of $\sim$20\% of O\textsubscript{2} remains trapped in the ASW matrix at the highest temperatures investigated.}
   {This work demonstrates that the surface diffusion of IR-inactive molecules can be experimentally quantified using mass spectrometry. Our findings show that O\textsubscript{2} exhibits a low diffusion barrier, indicating high mobility in interstellar water ices. Moreover, we suggest these water ices likely retain a residual fraction of hypervolatiles entrapped within their structure.}

   \keywords{astrochemistry $-$ diffusion $-$ ISM: molecules $-$ methods: laboratory: solid state $-$ molecular processes}

  \maketitle

\section{Introduction}
In the interstellar medium (ISM), molecular clouds are composed of 99\% gas by mass, where reactions are unlikely to occur because molecular density rarely exceeds 100 molecules per cm\textsuperscript{$-$3} \citep{lammer2009, gent2013}, and 1\% dust \citep{linnartz2015}. In the coldest regions, where temperatures drop below 20 K, gaseous atoms and molecules freeze out onto dust grains forming ice mantles where complex chemical processes take place \citep{boogert2015, mcclure2023}. This solid phase provides reactive sites for the formation of simple and complex organics, owing to its higher reactants concentration and its role as a third body capable of dissipating excess energy from the reaction thus stabilizing the products \citep{van2014, linnartz2015, cuppen2017, oberg2021}.

Even though this rich ice chemistry can proceed through non-diffusive mechanisms, it is largely facilitated by diffusion \citep{ligterink2025}, a thermally activated process that refers to the mobility of molecules enabling them to meet and react. At the prestellar cloud stages, when the temperatures are very low, small hypervolatile species like H atoms can diffuse and initiate reactions (e.g. hydrogenation) with already adsorbed atoms and molecules such as atomic and molecular oxygen (O, O\textsubscript{2}) \citep{ioppolo2010} or carbon monoxide (CO). These reaction pathways mainly lead to the formation of amorphous solid water (H\textsubscript{2}O) \citep{cuppen2010}, the major component of interstellar ices \citep{dartois2005}. In addition, under these conditions, a fraction of O\textsubscript{2} molecules might not react and are expected to be mixed within these water-rich ice layers \citep{boogert2015}. Moreover, when a protostar forms and heats its surroundings, the ice undergoes thermal processing that restructures its morphology and modifies diffusion properties; as diffusion rates increase, molecular complexity rises \citep{minissale2013}. Indeed, with heating of the ice additional diffusion of the small organics through the water ice matrix leads to the formation of more complex molecules \citep{ligterink2025}, such as methanol (CH\textsubscript{3}OH), formaldehyde (CH\textsubscript{2}O) \citep{rimola2014,fuchs2009}, or formic acid (CH\textsubscript{2}O\textsubscript{2}) \citep{inostroza2024}, constituting a vast reservoir of molecular diversity. The efficiency of ice chemistry is therefore heavily controlled by the O\textsubscript{2} and other reactants' diffusion. The warm-up episodes induce molecular desorption in the gas phase even though some species remain entrapped in the ice matrix. In the end, the exact balance between the competing processes of diffusion, entrapment, and desorption dictates the composition of the solid and gas phases during star formation \citep{minissale2016a}. 

At temperatures typically below 140 K, surface diffusion, governed by the Langmuir–Hinshelwood process, is the dominant mechanism and takes place on either an external surface or within internal pores and cracks of the ice \citep{mispelaer2013, lauck2015, ligterink2025}. This process is believed to occur by thermal hopping over energy barriers or by quantum tunneling, prevailing at very low temperatures (<10 K)  \citep{minissale2013}. On the other hand, at higher temperatures the ice gets more compact \citep{bossa2012} and bulk diffusion becomes the main mechanism. Modeled as a swapping process or as movement into interstitial spaces between molecules, bulk diffusion possesses higher energy barriers \citep{mispelaer2013, lauck2015}. 

Having access to diffusion parameters of various molecular species of icy astrochemical environments helps to understand and predict the chemical evolution of star-forming regions. Accurately determined diffusion coefficients, experimentally or theoretically, are essential to be input in astronomical models \citep{hasegawa1992, acharyya2022, dijkhuis2025}, but are often poorly constrained and might vary by orders of magnitude. Indeed, diffusion itself is challenging to study as diffusion barriers are usually inferred from diffusion-limited processes \citep{cuppen2024}. However, because only few species (e.g., CO \citep{lauck2015} or CH\textsubscript{4} \citep{mate2020}) display such behavior in a measurable way and instrumental sensitivity often imposes additional constraints, the set of molecules for which diffusion barriers can be extracted remains limited. To deal with this lack of data, modelers established various ratios (0.3–0.8) \citep{tielens1982, garrod2011, chang2012} to derive diffusion coefficients from the binding energies, a parameter easier to determine, although this introduces large uncertainties in their models \citep{cuppen2024, furuya2022}. 

Several previous experimental studies using different techniques under ultrahigh-vacuum conditions have therefore aimed to accurately determine these diffusion coefficients in complement to theoretical investigations. A common way to extract those of stable species is by using infrared (IR) spectroscopy on layered or mixed ices, allowing the monitoring of target species diffusion during an isothermal experiment. For instance, data are largely available for the surface diffusion of CO in ASW ice at low temperature ($\leq$50 K). \citet{oberg2009} and \citet{karssemeijer2013} determined a similar diffusion energy barrier for CO of 26 ± 9 meV and 26 ± 15 meV respectively, while \citet{mispelaer2013} reported a lower energy barrier with a value of 10 ± 15 meV, and \citet{lauck2015} established a diffusion energy barrier of 14 ± 1 meV. These results not only show typical values for diffusion energy barriers but also the importance of taking into account uncertainties and variabilities between results. Studying the diffusion of unstable species such as radicals is more challenging. Some techniques rely on following diffusion-limited processes such as the formation of specific products. Considering the diffusion of O atoms, one way to extract their diffusion coefficients is by monitoring O\textsubscript{2} and O\textsubscript{3} formation using mass spectrometry \citep{minissale2016b}. Sometimes, studying the diffusion of stable molecules may also prove challenging. For instance, homonuclear diatomic species such as O\textsubscript{2} cannot be as easily monitored as CO because of their infrared-inactive nature. Consequently, O\textsubscript{2} remains poorly described in models, despite being of clear astrochemical interest: it is expected to be a major reservoir of elemental oxygen in dense molecular clouds \citep{herbst1973} and, once adsorbed on icy dust grains, its diffusion is likely to play a major role in increasing the molecular complexity of the ice phase \citep{taquet2016, ligterink2025}. To the best of our knowledge, O\textsubscript{2} diffusion in ASW has only been experimentally studied once, indirectly, by probing its impact on the water O–H stretching mode through IR spectroscopy \citep{he2018}. 

\begin{figure}
  \resizebox{\hsize}{!}{\includegraphics{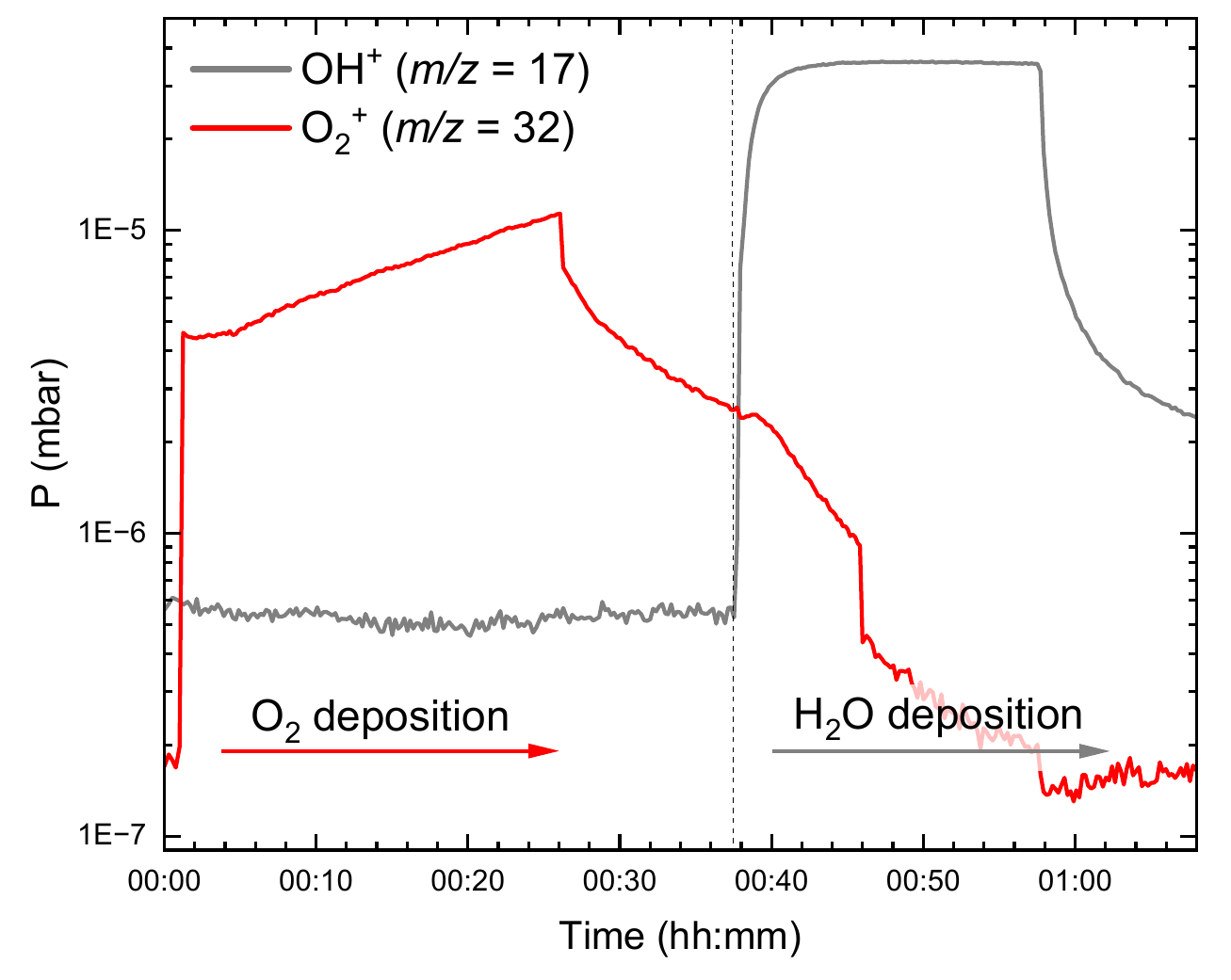}}
  \caption{QMS signals recorded during the consecutive deposition of O\textsubscript{2}  at $m/z$ = 32 and H\textsubscript{2}O at $m/z$ = 17.}
  \label{Fig. 1}
\end{figure}

Here, we present an alternative approach in which O\textsubscript{2} kinetics, in initially layered O\textsubscript{2}–H\textsubscript{2}O interstellar ice analogs, is directly monitored using mass spectrometry upon surface diffusion followed by instant desorption into the gas phase. This study aims to determine the diffusion energy barrier of O\textsubscript{2} in ASW by measuring its Fickian diffusion coefficients (D) at different temperatures (35 K, 40 K and 45 K) ensuring diffusion is the limiting process compared to desorption. The D values are expected to be temperature-dependent and follow an Arrhenius trend. This technique opens up possibilities for the determination of the diffusion coefficients of other infrared-inactive species such as N\textsubscript{2} or noble gases. The entrapment efficiency of O\textsubscript{2} in H\textsubscript{2}O as a function of temperature has also been investigated in the present work. 

Sect.~\ref{sec:2} describes the experimental setup and procedures, as well as the details for quadrupole mass spectrometer (QMS) monitoring and data analysis using a Fickian equation. It also motivates the choice of the Fickian equation for fitting the experimental data and extracting the diffusion parameters. Sect.~\ref{sec:3} reports the derived diffusion coefficients of O\textsubscript{2} and \textsuperscript{18}O\textsubscript{2}, and discusses their dependence on temperature and ice thickness. The extracted diffusion energy barrier and entrapment efficiencies in H\textsubscript{2}O ice are also presented in this section. The astrophysical implications of the newly obtained values are highlighted in Sect.~\ref{sec:4}, before concluding with the main findings in Sect.~\ref{sec:5}.

\section{Methods}

\label{sec:2}

\subsection{Experimental setup}

The experiments are conducted in the ultra-high vacuum (UHV) setup $\mathrm{SURFRESIDE^{3}}$, which has already been described in detail by \citet{ioppolo2013} and \citet{santosthesis}. It consists of a main chamber reaching a typical base pressure of $\sim$3 $\times 10^{- 9}$ mbar at room temperature and presenting, at its center, a gold-plated copper substrate mounted on the tip of a closed-cycle helium cryostat. The substrate temperature can be adjusted between 9 K and 450 K using resistive heaters and is controlled with a precision of $\pm$ 0.5 K using two silicon diode sensors. Gases of O\textsubscript{2} (Linde, purity 99.9\%) and \textsuperscript{18}O\textsubscript{2} (Campro, 97\% atom \textsuperscript{18}O) as well as vapors from deionized H\textsubscript{2}O are prepared individually in a separate deposition line. The H\textsubscript{2}O sample was purified by several freeze-pump-thaw cycles. The species are then introduced sequentially into the main chamber through a high-precision leak valve facing the substrate, allowing their direct deposition on the cold substrate kept at 10 K. Fourier-transform reflection-absorption infrared spectroscopy (RAIRS) is employed to monitor the ice growth and the H\textsubscript{2}O structure in real time by recording spectra with 1 cm\textsuperscript{$-$1}  resolution in the 700–4000 cm\textsuperscript{$-$1} range. The gas-phase composition containing desorbing species is sampled using a QMS.
\begin{figure}
  \resizebox{\hsize}{!}{\includegraphics{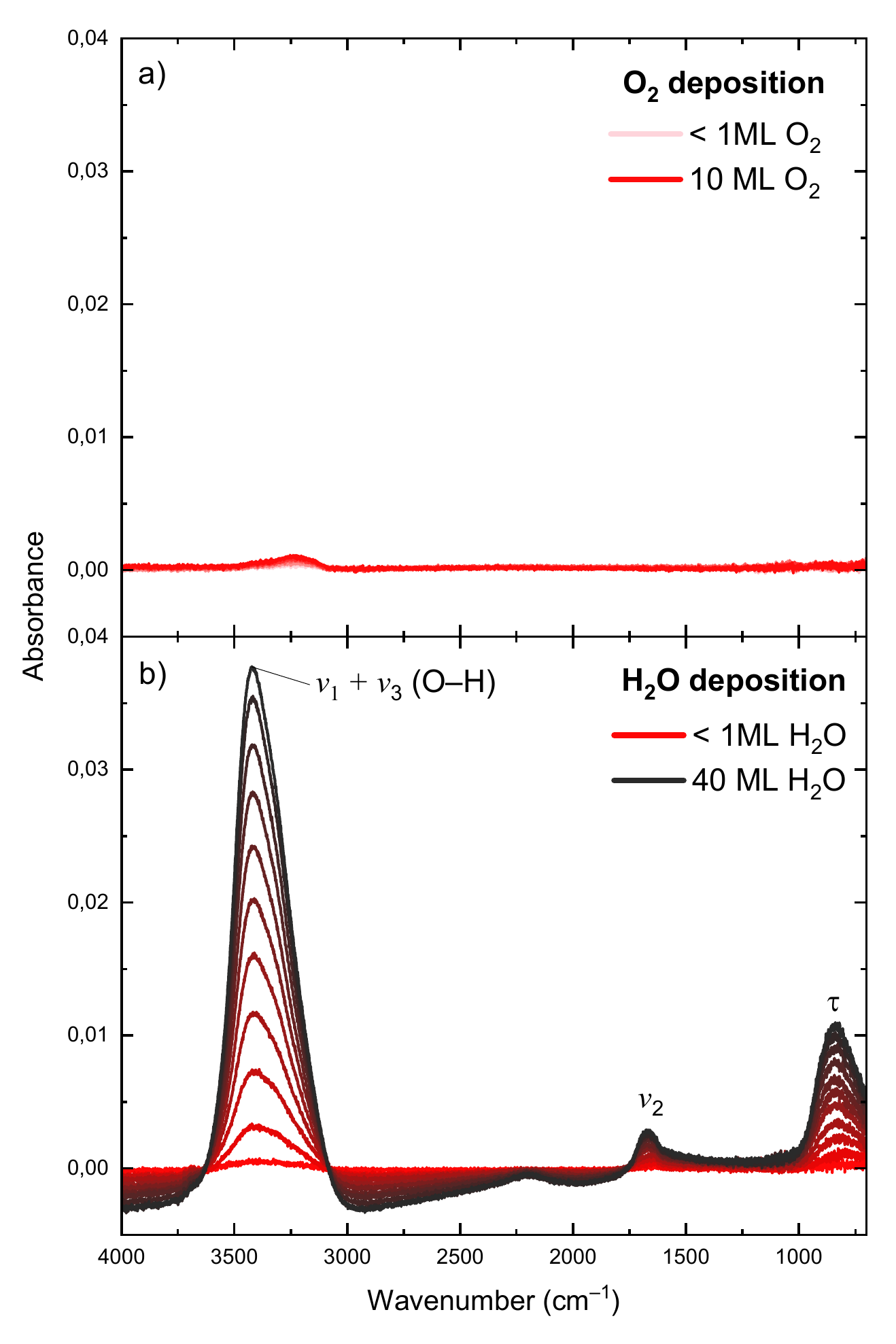}}
  \caption{Infrared spectra recorded for experiment 4a during the consecutive deposition of O\textsubscript{2} (a) and H\textsubscript{2}O (b), over the 400 to 4000 cm\textsuperscript{$-$1} range.
  \textit{Top} (a): No infrared band is observed, the small feature around 3300 cm\textsuperscript{$-$1} corresponds to saturation of the water feature from crystallized water in the MCT detector.
  \textit{Bottom} (b): $\nu_{1} + \nu_{3}$ corresponds to the O$-$H stretching mode, $\nu_{2}$ to the H\textsubscript{2}O bending mode, and $\tau$ to the water libration mode. The increase in the intensity of these bands over time reflects the growth of the H\textsubscript{2}O ice thickness.}
  \label{Fig. 2}
\end{figure}

\begin{table}
\caption{Parameters used to quantify the species' abundances with the QMS.}                 
\label{table:1}    
\centering                        
\begin{tabular}{c c c c}      
\hline\hline               
Species & $\sigma^{+}$(\AA\textsuperscript{2}) & $F_{\mathrm{F}}(m)$ \tablefootmark{d} & $S(m/z)$ \tablefootmark{e} \\         
\hline                      
   O\textsubscript{2} & 1.69 \tablefootmark{b} & 0.82 & 0.30 \\    
   \textsuperscript{18}O\textsubscript{2} \tablefootmark{a} & 1.69 \tablefootmark{b} & 0.82 & 0.26 \\
   H\textsubscript{2}O & 1.27 \tablefootmark{c} & 0.17 & 0.58\\
\hline                                  
\end{tabular}
\tablefoot{\tablefoottext{a}{$\sigma^{+}$ and $F_{\mathrm{F}}(m/z)$ for \textsuperscript{18}O\textsubscript{2} are assumed to be the same as for O\textsubscript{2};}
\tablefoottext{b}{\citet{krishnakumar1992};}
\tablefoottext{c}{\citet{itikawa2005};}
\tablefoottext{d}{NIST;}
\tablefoottext{e}{\citet{chuang2018}.}
}

\end{table}

\subsection{Experimental procedure}

\label{sec:2.2}

Configuration of the studied ice sample is a bilayer structure composed of a thin O\textsubscript{2} layer completely covered by H\textsubscript{2}O. After cooling down the substrate to 10 K, 10 monolayers (ML) of O\textsubscript{2} are deposited at a flow rate of $6.9 \times 10^{12}$ molecules cm\textsuperscript{$-$2} min\textsuperscript{$-$1}, as determined from previous leak-valve calibration steps. Subsequently, deposition of H\textsubscript{2}O is performed with variable thicknesses (40, 60, 80 ML) at a flow rate of $2.4 \times 10^{13}$ molecules cm\textsuperscript{$-$2} min\textsuperscript{$-$1}, forming an amorphous and porous water ice layer under the cold deposition conditions. The gas composition within the chamber is monitored throughout deposition using the QMS (Fig.~\ref{Fig. 1}). The deposited ice sample is further heated with a ramp of 5 K min\textsuperscript{$-$1} to designated temperatures at 25, 30, 35, 40, and 45 K and is then held isothermally. This entire four-hour (three-hour for Exp 1) process constitutes the isothermal phase. The thermal process confers energy to the O\textsubscript{2} molecules, enabling them to migrate through the H\textsubscript{2}O layer toward the surface. The desorbing species are ionized by 70 eV electron impact and monitored with the QMS. After that, the ice undergoes temperature-programmed desorption (TPD), a linear heating to 300 K with a ramp of 5 K min\textsuperscript{$-$1}. The desorption profile during that phase is used to quantify entrapment efficiencies of the volatiles still present in the ice.

The surface coverage is estimated using two approaches. For H\textsubscript{2}O, the ice growth is monitored in situ during deposition using RAIRS (Fig.~\ref{Fig. 2}) and the final column density is obtained using the IR integrated absorbance ($\int Abs(\nu)d\nu$) of the OH stretching band at $\sim$3352 cm\textsuperscript{$-$1} and converting it to absolute abundance with a modified Beer-Lambert law:
 \begin{equation}
    \label{eq1}
      N_{\mathrm{x}}  =  \ln10 \cdot \frac{\int Abs(\nu)d\nu}{A^{'}(X)} \,,
   \end{equation}
with $N_{\mathrm{x}}$ being the species' column density in molecules cm\textsuperscript{$-$2} and $A^{'}(X)$ its absorption band strength in cm molecule\textsuperscript{$-$1}. Based on values reported by \citet{gerakines1995}, the H\textsubscript{2}O band strength has been taken as $A^{'}$(H\textsubscript{2}O)\textsubscript{O-H str }= (2.0 $\pm$ 0.4) $ \times $ 10\textsuperscript{$-$16} cm molecule\textsuperscript{$-$1} and has been corrected using a setup-specific conversion factor from transmission to reflection mode of 3.2, measured by \citet{santos2023} with the same experimental setup. The final $N_{\mathrm{{H\textsubscript{2}O}}}$ is estimated with a relative uncertainty of 20\% mainly coming from the band strength. 

\begin{table*}[h!]
\caption{Experimental details for all initially layered O\textsubscript{2}$-$H\textsubscript{2}O, \textsuperscript{18}O\textsubscript{2}$-$H\textsubscript{2}O, and pure H\textsubscript{2}O ice experiments.}                 
\label{table:2}    
\centering                        
\begin{tabular}{c c c c c}      
\hline\hline               
Experiment & Species & O\textsubscript{2}/H\textsubscript{2}O (ML) & Isothermal T (K) & Isothermal duration (h) \\         
\hline                      
   0 & H\textsubscript{2}O & 0/42 & 40 & 4\\    
\hline       
   1 & O\textsubscript{2}$-$H\textsubscript{2}O & 23/41 & 25 & 3 \\
   2 & O\textsubscript{2}$-$H\textsubscript{2}O & 11/40 & 30 & 4 \\
   3 & O\textsubscript{2}$-$H\textsubscript{2}O & 11/41 & 35 & 4 \\
   4 & O\textsubscript{2}$-$H\textsubscript{2}O & 10/42 & 40 & 4 \\
   5 & O\textsubscript{2}$-$H\textsubscript{2}O & 12/39 & 45 & 4 \\
\hline
   6 & O\textsubscript{2}$-$H\textsubscript{2}O & 15/60 & 40 & 4 \\
   7 & O\textsubscript{2}$-$H\textsubscript{2}O & 7/79 & 40 & 4 \\
\hline
   8 & \textsuperscript{18}O\textsubscript{2}$-$H\textsubscript{2}O & 5/40 & 35 & 4 \\
   9 & \textsuperscript{18}O\textsubscript{2}$-$H\textsubscript{2}O & 5/40 & 40 & 4 \\
   10 & \textsuperscript{18}O\textsubscript{2}$-$H\textsubscript{2}O & 5/41 & 45 & 4 \\
\hline
   4a & O\textsubscript{2}$-$H\textsubscript{2}O & 11/42 & 40 & 4 \\
   4b & O\textsubscript{2}$-$H\textsubscript{2}O & 16/39 & 40 & 4 \\
   4c & O\textsubscript{2}$-$H\textsubscript{2}O & 16/42 & 40 & 4 \\
   4d & O\textsubscript{2}$-$H\textsubscript{2}O & 11/43 & 40 & 4 \\
   4e & O\textsubscript{2}$-$H\textsubscript{2}O & 19/41 & 40 & 30 \\
\hline
\end{tabular}
\end{table*}

For infrared-inactive molecules such as O\textsubscript{2}, the growth needs to be estimated differently. To guide the deposition, the target O\textsubscript{2} ice coverage is predicted beforehand according to the Langmuir estimation by following the main chamber pressure\footnote{Assuming that 1 Langmuir, corresponding to an exposure of 10\textsuperscript{–6} Torr during one second, leads to the surface coverage of 1 ML.}. The final column density ($N_{\mathrm{x}}$) can be derived from the QMS data by using the following expression \citep{martin2015}:
\begin{equation}
    \label{eq2}
      N_{\mathrm{x}}  = k_{\mathrm{QMS}} \cdot \frac{A_{\mathrm{exp}}(m/z)}{\sigma^{+}_{\mathrm{x}} \cdot I_{\mathrm{F}}(z) \cdot F_{\mathrm{F}}(m) \cdot S(m/z)} \,,
   \end{equation}
where $k_{\mathrm{QMS}}$ is the proportionality constant of the apparatus, $A_{\mathrm{exp}}(m/z)$ is the integrated desorption signal of a given mass fragment, $F_{\mathrm{F}}(m)$ is its fragmentation fraction, and $S(m/z)$ is the corresponding sensitivity of the QMS. The sensitivity factor is setup-dependent and the values were extracted from the noble gas calibration experiment previously done by \citet{chuang2018}. Furthermore, $\sigma^{+}_{\mathrm{x}}$ stands for the molecule’s electronic ionization cross-section in \AA\textsuperscript{2} and $I_{\mathrm{F}}(z)$ is the fraction of ions with charge z (here we use $I_{\mathrm{F}}(z)$ = 1). For each species studied in this work, employed parameters are summarized in Table~\ref{table:1}. On the other hand, $k_{\mathrm{QMS}}$ is measured considering that it is independent of the species and assuming a relatively constant pumping speed for all explored species. In this study, we have $k_{\mathrm{QMS}}$ = (7.89 $\pm$ 1.76) $\times$ 10\textsuperscript{21} a.u. (see Appendix ~\ref{app:A}), with a 22\% uncertainty considered as the main source of error in $N_{\mathrm{O_{\mathrm{2}}}}$ estimation. All the final thicknesses were expressed in monolayer units for ease of comparison across experiments using the convention that 1 ML corresponds to 10\textsuperscript{15} molecules cm\textsuperscript{$-$2}.

The ice thickness of each sample in terms of height ($h_{\mathrm{x}}$ in cm) is obtained thanks to the species’ column density ($N_{\mathrm{x}}$) using the expression:
\begin{equation}
    \label{eq3}
      h_{\mathrm{x}} = \frac{N_{\mathrm{x}}\cdot M_{\mathrm{x}}}{\rho_{\mathrm{x}}\cdot N_{a}} \,,
   \end{equation}
where $M_{\mathrm{x}}$ is the species’ molar mass in g mol\textsuperscript{$-$1}, $Na$ the Avogadro number in mol\textsuperscript{$-$1}, $\rho_{\mathrm{x}}$ the density considering $\rho_{\mathrm{O_{\mathrm{2}}}}$ = 1.54 g cm\textsuperscript{$-$3}, $\rho_{\mathrm{^{\mathrm{18}}O_{\mathrm{2}}}}$ = 1.73 g cm\textsuperscript{$-$3} \citep{fulvio2009}, and $\rho_{\mathrm{H_{\mathrm{2}}O}}$ = 0.60–0.66 g cm\textsuperscript{$-$3} depending on the experimental temperature \citep{brown1996}. The total thickness $h$ corresponding to the sum of the oxygen ($h_{\mathrm{O_{\mathrm{2}}}}$) and water ($h_{\mathrm{H_{\mathrm{2}}O}}$) heights is estimated with an uncertainty of 17\% and 18\% for O\textsubscript{2}–H\textsubscript{2}O and \textsuperscript{18}O\textsubscript{2}–H\textsubscript{2}O samples, respectively, based on error propagation from the uncertainties on the column densities.

All experiments performed in this study are summarized in Table~\ref{table:2}. It includes control experiments with \textsuperscript{18}O\textsubscript{2}, pure H\textsubscript{2}O ice, or longer isothermal phase, and mentions repeated experiments conducted to assess reproducibility, giving an experimental uncertainty of about 14\% on the QMS signal. 

\subsection{The analysis}

\subsubsection{Determination of diffusion coefficients and diffusion energy barrier}

\label{sec:2.3.1}

At a given temperature, the concentration profile of a diffusing species, $n(z,t)$, as a function of time, $t$, and position $z$, follows Fick’s second law of diffusion in one dimension: 
\begin{equation}
    \label{eq4}
      \frac{\partial n(z,t)}{\partial t} = D(T) \cdot \frac{\partial^2 n(z,t)}{\partial z^{2}} \,,
   \end{equation}
Given the successful use of this approach in previous studies on CO diffusion in ASW \citep{mispelaer2013, karssemeijer2013, lauck2015}, this Fickian law, described by the temperature-dependent diffusion constant $D$, was chosen to model O\textsubscript{2} diffusion in ASW, assuming it is the dominant mechanism. In the present study, the hypothesis of a homogeneous pre-mixed O\textsubscript{2}–H\textsubscript{2}O ice when starting the diffusion analysis (see Sect.~\ref{sec:3.1.1}) imposes the initial condition $n(z,0) = n_{\mathrm{0}}$ for $0 < z < h$, where $h$ denotes the total ice thickness. In addition, we set $n(h,t) = 0$, reflecting the immediate desorption of O\textsubscript{2} upon reaching the ice surface (see Sect.~\ref{sec:3.1.1}), and $\frac{\partial n(0,t)}{\partial z} = 0 $ indicating the absence of any O\textsubscript{2} desorption flux from the bottom of the ice. Using these boundary conditions, the resulting solution $n(z,t)$ follows the formulation described by \citet{karssemeijer2013}:
\begin{equation}
    \label{eq5}
      n(z,t) = \sum_{i=0}^{\infty} \frac{2n_{0}(- 1)^{i}}{\mu_{i}\cdot h} \cdot \cos(\mu_{i}\cdot z)\cdot \exp(- \mu_{i}^{2}\cdot D \cdot t) \,,
   \end{equation}
with $ \mu_{i} = \frac{(2i + 1)\cdot \pi}{2h}$, and $D$ being the diffusion coefficient. By integrating this expression over $z$ between $0$ and $h$, the column density of diffusing O\textsubscript{2} can be expressed and subsequently converted into QMS signal areas, $A(t)$, following the approach of \citet{karssemeijer2013}. However, unlike IR band areas which probe the remaining molecules in the sample, this study relies on QMS monitoring that records the molecules desorbing over time. Therefore, experimental data accounting for the amount of O\textsubscript{2} molecules diffusing within the ice are defined by $A(t) = A_{\mathrm{0}} - A_{\mathrm{exp}}(t)$, where $A_{\mathrm{0}}$ is the total integrated area of desorbing O\textsubscript{2} during the isothermal phase and $A_{\mathrm{exp}}(t)$ is the integrated desorption signal of O\textsubscript{2} during the same phase. The data are taken from the maximum of the isothermal desorption peak until the plateau reached just before the start of the TPD, and are fitted using the following final expression: 
\begin{equation}
    \label{eq6}
      A(t) = s + \sum_{i=0}^{\infty} \frac{2(A_{\mathrm{0}} - s)}{\mu_{i}^{2} \cdot h^{2}} \cdot \exp(-\mu_{i}^{2} \cdot D \cdot t) \,,
   \end{equation}
where $s$ is an experimental offset introduced to reproduce the experimental data, expected to remain close to zero in our case\footnote{According to Eq.~\ref{eq6}, $A(t)$ converges to $s$ at the end of the isothermal phase. However, $A(t)$ should also converge to 0 at that point, since $A_{\mathrm{exp}}(t)$ gradually achieves $A_{0}$ value, implying that $s$ = 0 in theory.} (see Table ~\ref{table:3}). The fixed parameters $A_{\mathrm{0}}$ and $h$ are obtained from spectroscopic measurements, while $D$ and $s$ are treated as free parameters and derived from the fit together with their associated uncertainties.

Considering that diffusion is an Arrhenius-type process, we use subsequently the diffusion coefficients $D$ obtained at different isothermal temperatures to derive the diffusion energy barrier $E_{\mathrm{D}}$ using the following Arrhenius equation:
\begin{equation}
    \label{eq7}
      D(T) = D_{\mathrm{0}} \cdot \exp(\frac{- E_{\mathrm{D}}}{k_{B}\cdot T}) \,,
   \end{equation}
where $D_{\mathrm{0}}$ is the pre-exponential factor in cm\textsuperscript{2} s\textsuperscript{$-$1}, $E_{\mathrm{D}}$ is the energy barrier in J, $k_{B}$ is the Boltzmann constant in J K\textsuperscript{$-$1}, and T is the temperature in K. The energy barrier for diffusion and the pre-exponential factor are extracted by linearly fitting $\ln(D)$ at different temperatures versus $\frac{1}{T}$. Uncertainty on $E_{\mathrm{D}}$ is obtained by performing a Monte Carlo (MC) sampling analysis where the initial parameter space is explored using 10,000 simulated independent trials.

\subsubsection{Entrapment efficiency determination}

\label{sec:2.3.2}

Instead of diffusing across the ice and reaching the surface, some volatile molecules may remain trapped inside the ice matrix, even well above their sublimation temperature. Once substrate temperatures are high enough and H\textsubscript{2}O desorbs, it releases the remaining entrapped O\textsubscript{2} molecules into the gas phase. Entrapment efficiencies are quantified as the ratio between the integrated area of the late O\textsubscript{2} desorption peak in the TPD profile—delimited by the water desorption peak—and the integrated area of the entire O\textsubscript{2} desorption profile, from the beginning of the isothermal phase to the end of the TPD run. In other words, we are probing the amount of O\textsubscript{2} sublimating together with water, relative to its initial abundance in the ice. 
 
Contamination of the water sample by atmospheric molecular oxygen stemming from the valve in its glass vial was revealed by the presence of a signal in the QMS corresponding to O\textsubscript{2} during water deposition (Fig.~\ref{Fig. B.1}) and confirmed by a control experiment with water only (experiment 0, see Table~\ref{table:2}). Contamination does not impact the diffusion quantification, which is the main focus of our study, as this process is independent of the origin of the O\textsubscript{2} molecules. However, it interferes with entrapment efficiencies as pre-existing atmospheric O\textsubscript{2} molecules already occupy a fraction of the sites in the water ice matrix, leading to a lower apparent entrapment efficiency. For each experiment, entrapment efficiencies are therefore corrected with respect to the initial contamination by using a correlation factor obtained from repeated experiments (Fig.~\ref{Fig. B.2}). 
 
Entrapment efficiencies are determined at different temperatures from experiments with fixed total coverage to remove the influence of ice thickness. The uncertainty, estimated at 13\%, corresponds to the standard deviation of corrected efficiencies from repeated experiments. 
\begin{figure}
  \resizebox{\hsize}{!}{\includegraphics{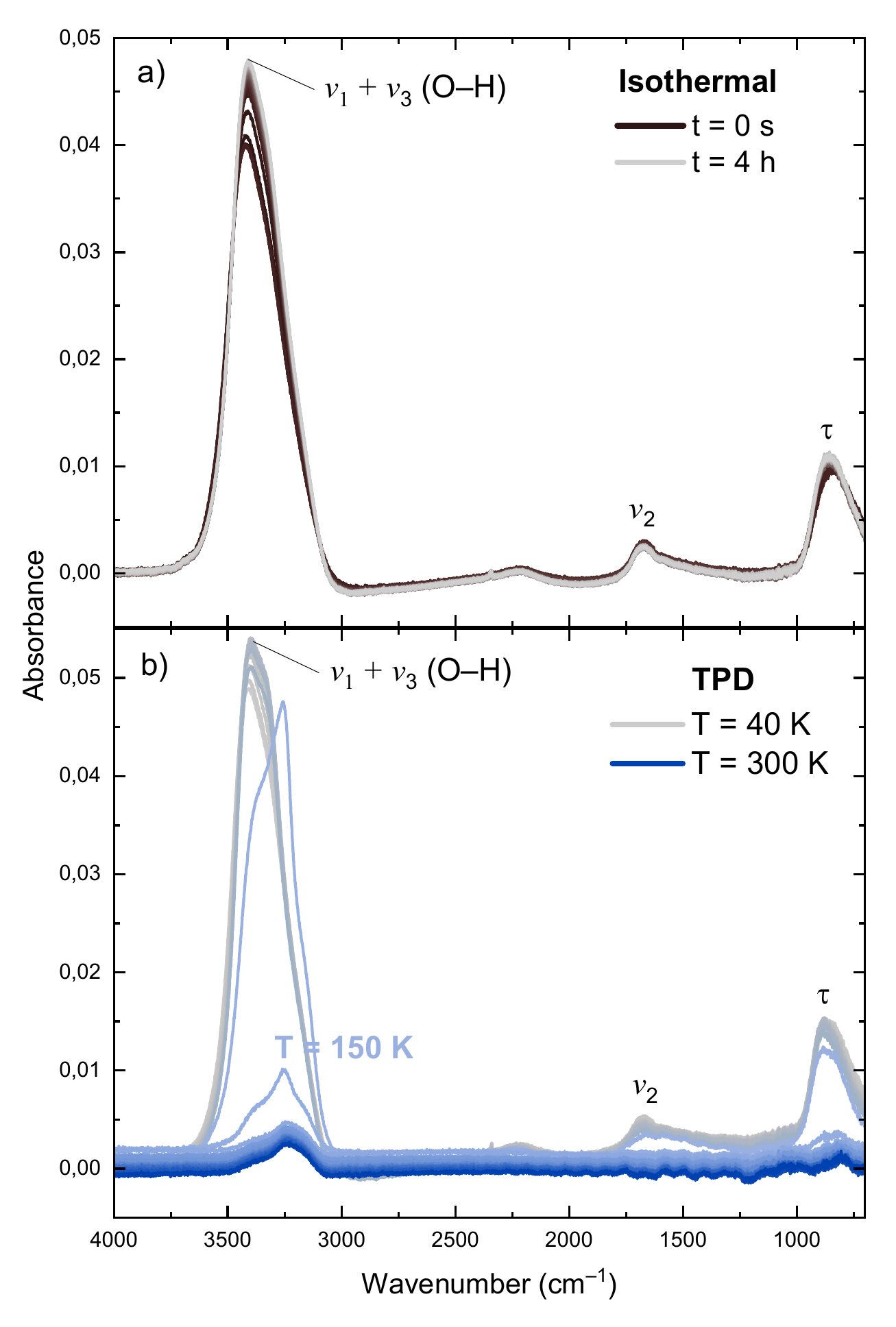}}
  \caption{Infrared spectra recorded for experiment 4a during the isothermal (a) and TPD (b) phases, over the 400 to 4000 cm\textsuperscript{$-$1} range. \textit{Top} (a): spectra were recorded every five minutes for four hours. \textit{Bottom} (b): spectra were recorded every two minutes during the heating from the isothermal temperature (40 K) to 300 K. At T = 150 K, the abrupt decrease of all characteristic H\textsubscript{2}O bands is attributed to water desorption.}
  \label{Fig. 3}
\end{figure}

\section{Results and discussion}

\label{sec:3}

\subsection{Diffusion in ASW}

Diffusion is quantified during the isothermal phase, when O\textsubscript{2} molecules desorb from a static water matrix. Figure~\ref{Fig. 3}(a) shows a typical infrared spectra (Exp. 4a) taken every 5 minutes during the four-hour isothermal phase presenting the characteristic infrared bands ($\nu_{1}$ + $\nu_{3}$ , $\nu_{2}$, and $\tau$) of the ASW ice while the O\textsubscript{2} mixed with H\textsubscript{2}O stays IR featureless, as mentioned in Fig.~\ref{Fig. 2}. It is important to note that a slight increase is observed at the beginning, within 20 minutes, of the isothermal phase in the $\nu_{1}$ + $\nu_{3}$ band, corresponding to +16\% in column density, the $\nu_{2}$ bending mode remains similar throughout the whole process. The absence of correlation between these bands rules out the possibility of additional deposition of water at the beginning of the isothermal phase, but rather indicates a slight reorganization of the ice while reaching the set temperature. After about 20 minutes, the water ice stabilizes and the IR spectra remain constant. 

\begin{figure*}[h!]
  \resizebox{\hsize}{!}{\includegraphics{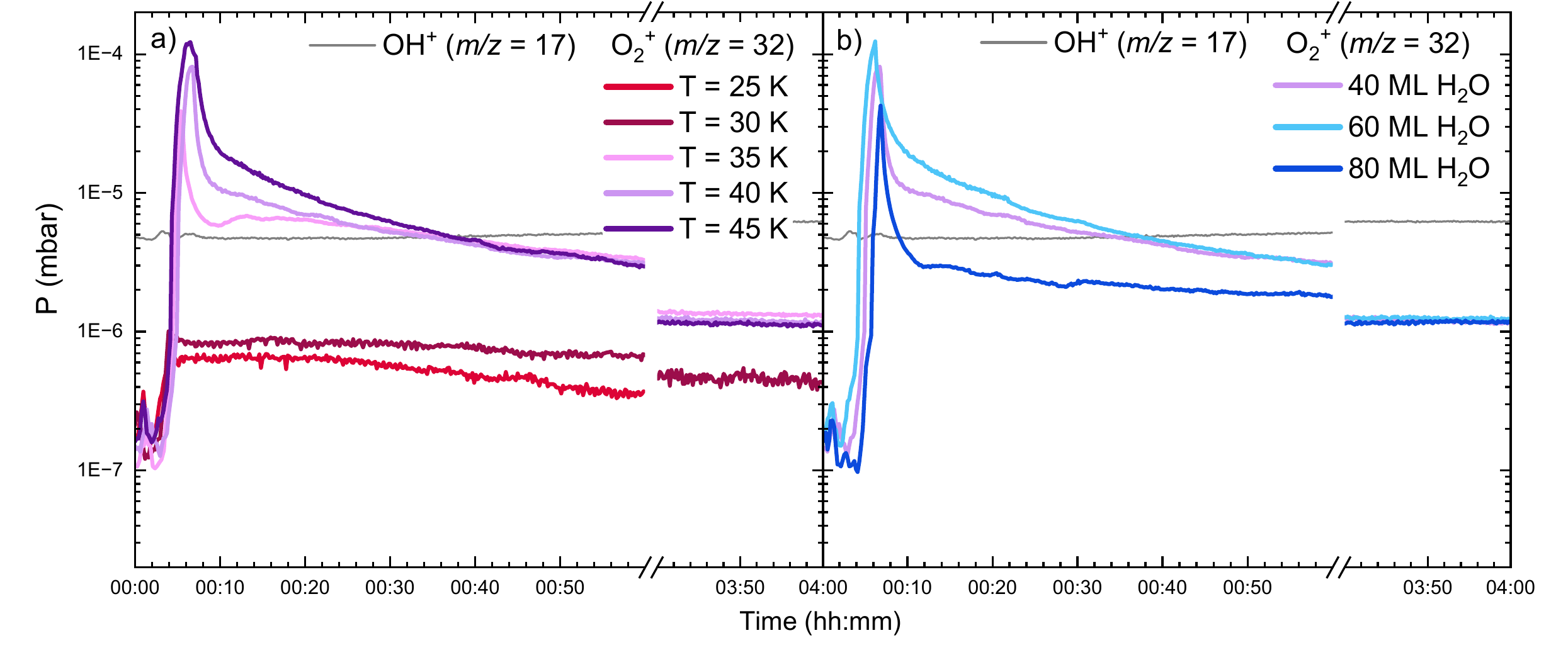}}
  \caption{QMS signals recorded during the four-hour isothermal phase at $m/z$ = 17 (OH\textsuperscript{+}) and $m/z $= 32 (O\textsubscript{2}\textsuperscript{+}). \textit{Left} (a): experiments 1 to 5 at different isothermal temperatures (25 K to 45 K with 40 ML of H\textsubscript{2}O). \textit{Right} (b): experiments 4, 6, and 7 with their different water ice thicknesses (40 ML to 80 ML at 40 K). Since the signal at $m/z$ = 17 was identical for every experiment, only the water profile of experiment 4 (T = 40 K; 40 ML H\textsubscript{2}O) is shown for clarity.}
  \label{Fig. 4}
\end{figure*}

\subsubsection{Diffusion coefficients}

\label{sec:3.1.1}

Diffusion coefficients of O\textsubscript{2} in ASW were determined from the isothermal phase of the experiments during which molecules diffuse freely. To accurately probe surface diffusion, it is necessary to follow an experimental regime that ensures diffusion is the limiting process. Consequently, isothermal sequences need to be conducted with high enough temperatures to guarantee instantaneous desorption and for a long enough experimental time to prevent any surface residence time effects. O\textsubscript{2} diffusion is monitored with QMS using the mass-to-charge ($m/z$) ratio of 32; the dominant signal corresponding to O\textsubscript{2}\textsuperscript{+} molecular ions produced by electron-impact ionization at 70 eV of the desorbing O\textsubscript{2} molecules\footnote{https://webbook.nist.gov/}. On the other hand, the constant signal at $m/z$ = 17 (OH\textsuperscript{+}) attests to the absence of water desorption at the isothermal temperatures. Water is monitored through this channel rather than through the molecular (abundant) mass signal of $m/z$ = 18 (H\textsubscript{2}O\textsuperscript{+}) of the QMS detector saturation. Figure~\ref{Fig. 4} shows the QMS data from experiments 1 to 7 (Table.~\ref{table:2}), conducted at different isothermal temperatures (Fig.~\ref{Fig. 4}(a): 25, 30, 35, 40, and 45 K with 40 ML of H\textsubscript{2}O) and with different water thicknesses (Fig.~\ref{Fig. 4}(b): 40, 60 and 80 ML at 40 K). For every experiment, a small feature is observed during the first two minutes of the isothermal phase, corresponding to an artifact resulting from the intense desorption of H\textsubscript{2}. For temperatures above 35 K, we observe a sharp desorption peak upon reaching the set temperature, followed by an exponential decrease of the signal ending with a plateau when diffusion is at its minimum. This sharp feature corresponds to the immediate desorption of weakly bound O\textsubscript{2} molecules that have already diffused and accumulated at the surface until their desorption is no longer limited. It highlights that the ices are initially mixed when starting the monitoring of diffusion. However, no such feature is observed at 25 K and 30 K, as the conditions do not ensure immediate desorption. Thus, only experiments in the proper diffusion-dominated regime were retained for further analysis (experiments 3 to 7). Additional \textsuperscript{18}O\textsubscript{2} experiments at different temperatures (35 to 45 K) were carried out to confirm that the diffusion mechanism is independent of water sample contamination (see Appendix ~\ref{app:C}, Fig.~\ref{Fig. C.1} for further explanations). 

\begin{figure*}[h!]
  \resizebox{\hsize}{!}{\includegraphics{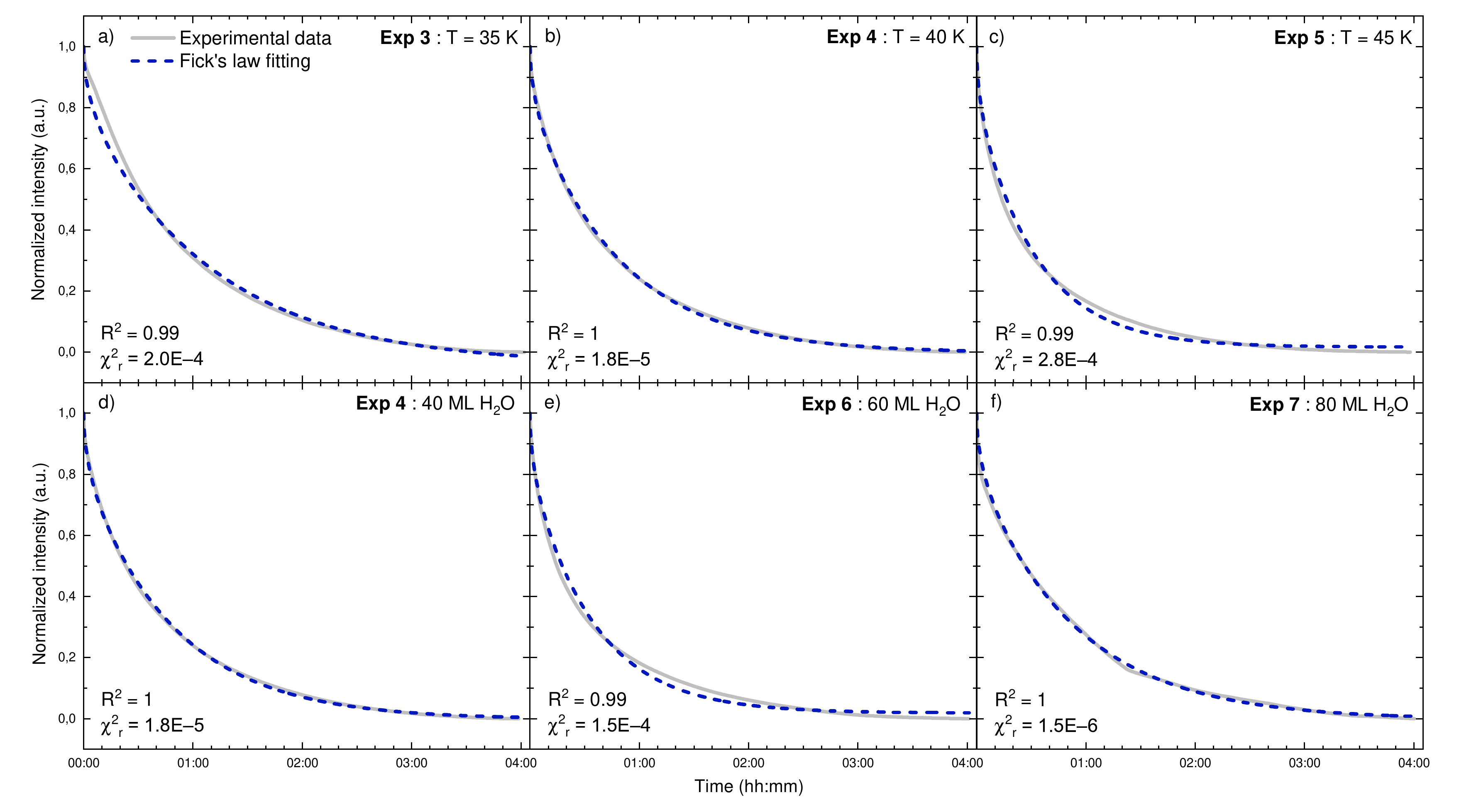}}
  \caption{Results of the fits to the Fickian model (blue) for all experimental data (grey) during the 4-hour isothermal phase. \textit{Top} (a)–(c): fits for experiments 3 to 5 conducted at 35 K, 40 K, and 45 K respectively, with a 40 ML water-ice matrix. \textit{Bottom} (d)–(f): fits for experiments 4, 6, 7 conducted  at 40 ML, 60 ML, and 80 ML respectively, for an isothermal temperature of 40 K.}
  \label{Fig. 5}
\end{figure*}

For every relevant experiment, the experimental decay curves ($A(\mathrm{t})$) ranging from the maximum of the desorption peak to the plateau, were fitted to Eq.~\ref{eq6}, derived from Fick’s second law of diffusion, where the fixed parameters $A_{0}$ and $h$ had to be determined (see Sect.~\ref{sec:2.2}). They correspond respectively to the total amount of desorbing O\textsubscript{2} during the isothermal phase and the total ice thickness, and are summarized in Table~\ref{table:3}. The resulting fits shown in Fig.~\ref{Fig. 5} demonstrate excellent agreement with the experimental data, as indicated by a coefficient of determination ($R$\textsuperscript{2}) greater than 0.99 and a $\chi$\textsuperscript{2} lower than 2.8 $\times$ 10\textsuperscript{$-$4}. Control \textsuperscript{18}O\textsubscript{2} experiments were also fitted to Eq.~\ref{eq6} and display good fit to the experimental data (see Appendix~\ref{app:C}). As mentioned in Sect.~\ref{sec:2.3.1}, diffusion coefficients $D$ and the offset $s$ were extracted from the fits and are listed in Table~\ref{table:3} alongside their respective uncertainties (see Table~\ref{table:C.1} for the isotopologue). 

\begin{table*}
\centering
\caption{Fixed parameters used for Fick’s diffusion law analysis of all initially layered O\textsubscript{2}–H\textsubscript{2}O experiments, along with their extracted diffusion rate.}                 
\label{table:3}    
\centering                        
\begin{tabular}{c c c c c}      
\hline\hline                    
Experiment & ${A}_{0}$ (10\textsuperscript{$-$7} mbar s\textsuperscript{$-$1}) \tablefootmark{a} & $h$ (10\textsuperscript{$-$6} cm) & $D$ (10\textsuperscript{$-$15} cm\textsuperscript{2} s\textsuperscript{$-$1}) & $s$ (10\textsuperscript{$-$8} mbar s\textsuperscript{$-$1}) \tablefootmark{b} \\         
\hline                      
   3 & 2.4 & 2.3 $\pm$ 0.4 & 0.5 $\pm$ 0.2 & $-$1.0 $\pm$ 2.9 \\    
   4 & 2.6 & 2.3 $\pm$ 0.4 & 0.7 $\pm$ 0.3 & $-$0.1 $\pm$ 2.0 \\
   5 & 3.4 & 2.2 $\pm$ 0.4 & 1.0 $\pm$ 0.3 & 0.5 $\pm$ 1.4 \\
   6 & 3.4 & 3.2 $\pm$ 0.6 & 2.0 $\pm$ 0.7 & 0.6 $\pm$ 2.1 \\
   7 & 0.8 & 3.9 $\pm$ 0.7 & 1.8 $\pm$ 3.5 & $-$0.1 $\pm$ 3.6 \\
\hline                                  
\end{tabular}
\tablefoot{\tablefoottext{a}{All $A\textsubscript{0}$ have an uncertainty of 1 $\times$ 10\textsuperscript{$-$10} mbar s\textsuperscript{$-$1};}
\tablefoottext{b}{The offset parameter $s$ is negligible in our study, being nearly two orders of magnitude smaller than $A$\textsubscript{0}.}
}

\end{table*}

\subsubsection{Dependence on temperature}

Experiments 3 to 5 (at 35 K, 40 K, and 45 K) were used to investigate the effect of temperature on diffusion efficiency. Figure~\ref{Fig. 4}(a) shows that for these experiments the desorption peak becomes more intense as the isothermal temperature increases. Indeed, when additional thermal energy is provided to the system, more O\textsubscript{2} molecules are able to instantly desorb upon reaching the set temperature \citep{minissale2022}. 

The measured diffusion coefficients show a clear temperature dependence as their value doubles from (0.5 $\pm$ 0.2) to (1.0 $\pm$ 0.3) $\times$ 10\textsuperscript{$-$15} cm\textsuperscript{2} s\textsuperscript{$-$1} when the temperature increases from 35 K to 45 K. This demonstrates that O\textsubscript{2} diffusion becomes more efficient as the ASW matrix is heated and microscopic mechanisms such as molecular hopping are activated. In addition, increasing the temperature induces morphological changes and pore rearrangements that further enhance molecular mobility.

\begin{figure}
  \resizebox{\hsize}{!}{\includegraphics{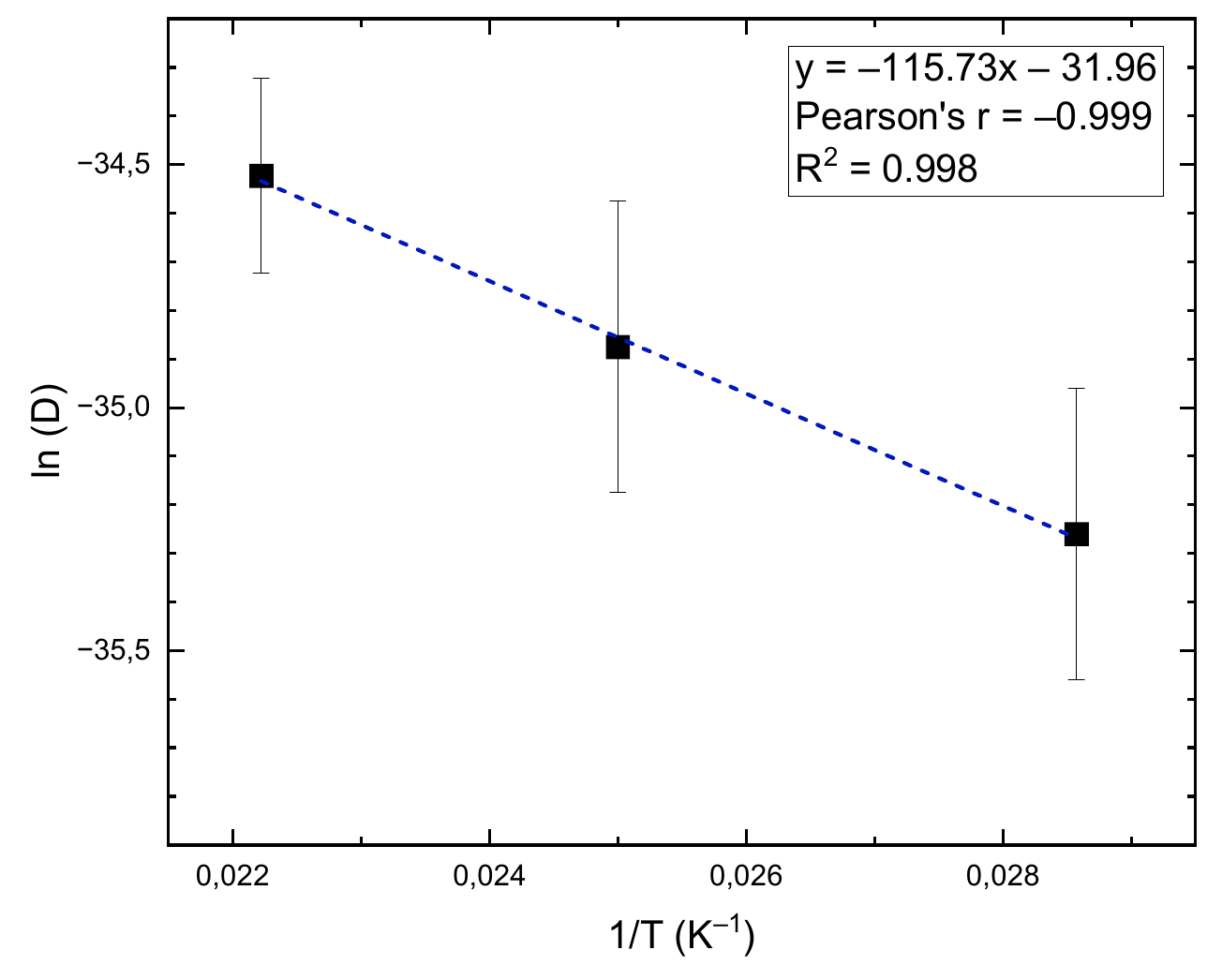}}
  \caption{Arrhenius plot showing the O\textsubscript{2} diffusion coefficients from experiment 3–5, obtained at different isothermal temperatures (35–45 K) with an ASW thickness of 40 ML. Error bars on $\ln(D)$ were determined through a MC analysis based on 10 000 independent trials.}
  \label{Fig. 6}
\end{figure}

As mentioned in Sect.~\ref{sec:2.3.1}, a linear fit of the logarithm of the diffusion coefficients versus $\frac{1}{\mathrm{T}}$ was performed (Fig.~\ref{Fig. 6}) resulting in a diffusion energy barrier of 10 $\pm$ 3 meV (116 $\pm$ 35 K). The fit shows excellent agreement with the experimental data, as indicated by the $R$\textsuperscript{2} value of 0.998 for the linear correlation, confirming that diffusion follows an Arrhenius-type behavior. The low value of the diffusion energy barrier explains why O\textsubscript{2} diffusion is rapidly activated upon heating of the ice. The value of the pre-exponential factor was also extracted from the linear fit, yielding $D_{0}$ = (2.3 $\pm$ 1.9) $\times$ 10\textsuperscript{$-$14} cm\textsuperscript{2} s\textsuperscript{$-$1}, which represents the theoretical upper limit of the diffusion coefficient as temperature approaches infinity.

\subsubsection{Dependence on water thickness}

Unlike temperature, the diffusion coefficients are expected to be independent of the water ice thickness as long as the initial conditions of the Fickian model remain the same. Experiments 4, 6, and 7 (with H\textsubscript{2}O ice layers of 40 ML, 60 ML, and 80 ML, respectively) were used to evaluate the effect of the water coverage on diffusion efficiency in our study. Whether considering the first desorption peak observed in the QMS signals (Fig.~\ref{Fig. 5}(b)) or the extracted diffusion coefficients summarized in Table~\ref{table:3}, no clear trend with the ice thickness is identified. Although $D$ increases from (0.7 $\pm$ 0.3) to (2.0 $\pm$ 0.7) $\times$ 10\textsuperscript{$-$15} cm\textsuperscript{2} s\textsuperscript{$-$1} when going from 40 ML to 60 ML of water ice, it then decreases to (1.8 $\pm$ 3.5) $\times$ 10\textsuperscript{$-$15} cm\textsuperscript{2} s\textsuperscript{$-$1}, when the ice is 80 ML thick. The latter’s large uncertainty does not allow for a clear interpretation. 

One hypothesis is that ASW undergoes rapid reorganization, even at low temperatures, as supported by a 16\% increase in the $\nu$\textsubscript{1} + $\nu$\textsubscript{3} band column density (Fig.~\ref{Fig. 3}) during the isothermal phase, and as described by \citet{slavicinska2024}. This rapid evolution of the ice morphology could lead to a distribution of pore and crack sizes in the ice throughout the isothermal phase. In that situation, the diffusion coefficient measured in a single experiment may in reality correspond to a distribution of diffusion coefficients, as mentioned by \citet{mispelaer2013}. For thicker ices, this phenomenon can become even more significant and could explain both the absence of a clear trend and the large uncertainties. Another factor, that makes this unexpected dependency with thickness difficult to interpret, is the uncertainty in the estimated O\textsubscript{2}/H\textsubscript{2}O ratios. Indeed, H\textsubscript{2}O contamination with atmospheric O\textsubscript{2} affects the effective O\textsubscript{2} coverage, and this phenomenon is exacerbated for thicker H\textsubscript{2}O ices. 

\begin{figure*}[h!]
  \resizebox{\hsize}{!}{\includegraphics{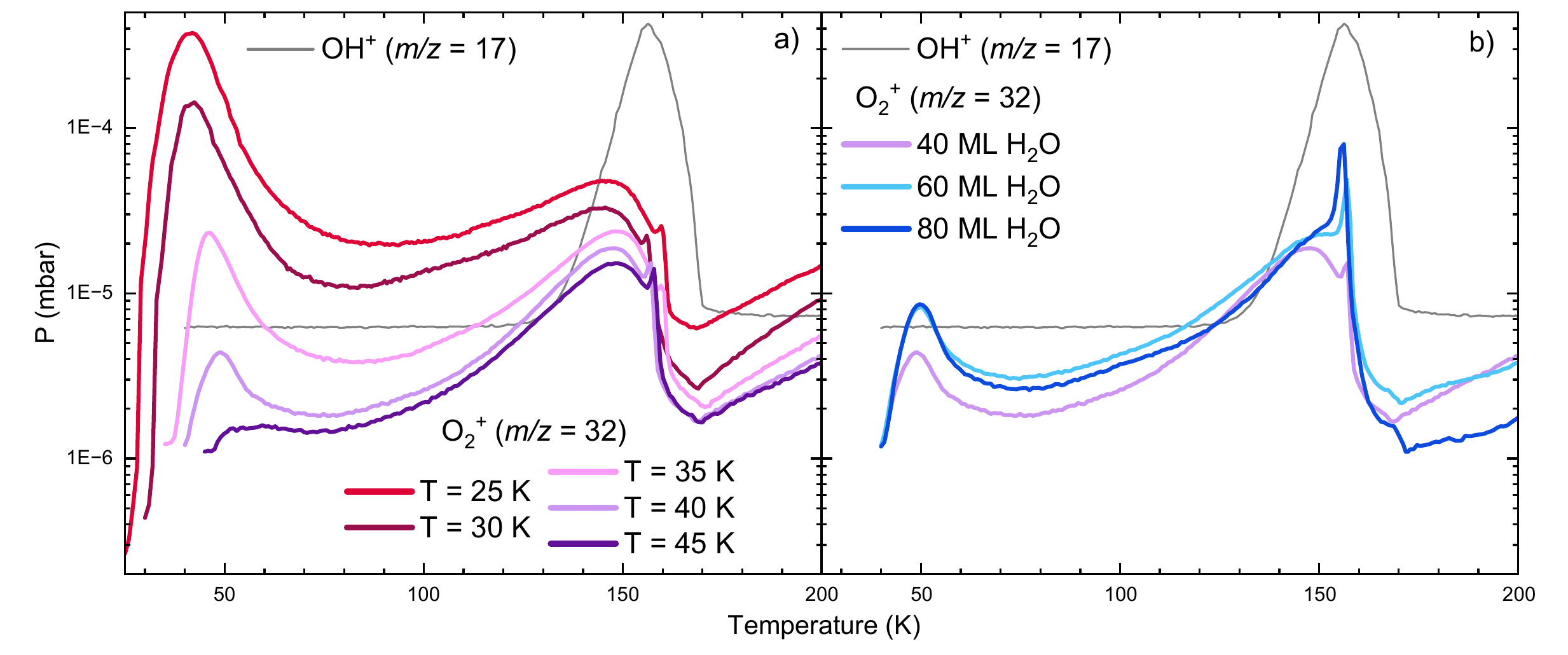}}
  \caption{QMS signals recorded during the TPD phase at $m/z $= 17 (OH\textsuperscript{+}) and $m/z$ = 32 (O\textsubscript{2}\textsuperscript{+}). \textit{Left} (a): experiments 1 to 5 at different isothermal temperatures (25 K to 45 K with 40 ML of H\textsubscript{2}O). \textit{Right} (b): experiments 4, 6, and 7 with their different water ice thicknesses (40 ML to 80 ML at 40 K). Since the signal at $m/z$ = 17, showing water desorption peaking at around 150 K, was identical for every experiment, only the water profile of experiment 4 (T = 40 K; 40 ML H\textsubscript{2}O) is shown for clarity. }
  \label{Fig. 7}
\end{figure*}

\subsection{Entrapment in ASW}

Entrapment of O\textsubscript{2} in ASW is studied by analyzing the TPD phase of our experiments. During this phase, infrared spectra were recorded every 2 minutes (Fig.~\ref{Fig. 3}(b)), showing two phenomena. Just before T = 150 K, the peak position of the stretching mode shifts to lower wavenumbers, corresponding to water crystallization upon heating of the ice \citep{mate2012}. Then, a sharp decrease in band intensity of all characteristic H\textsubscript{2}O IR modes, around 150 K, indicates water desorption. 

Figure~\ref{Fig. 7} presents the QMS data recorded from experiments 1 to 7 (Table~\ref{table:2}), conducted at different isothermal temperatures (Fig.~\ref{Fig. 7}(a): 25, 30, 35, 40, and 45 K with 40 ML of H\textsubscript{2}O) and with different water thicknesses (Fig.~\ref{Fig. 7}(b): 40, 60 and 80 ML at 40 K). At $m/z$ = 32, the TPD profiles reveal two desorption regimes. The first feature, around 50 K, corresponds to desorption of O\textsubscript{2} molecules that did not have sufficient time to diffuse to the surface during the four-hour isothermal phase. As expected, when the isothermal temperature increases (Fig.~\ref{Fig. 7}(a)), the area of this desorption peak decreases. Indeed, as more O\textsubscript{2} molecules diffuse and desorb during the isothermal phase, fewer remain to desorb at the beginning of the TPD. The second desorption feature, occurring around 140–160 K, corresponds to the entrapped O\textsubscript{2} molecules that desorb simultaneously with water. This feature increases with the water ice thickness (Fig.~\ref{Fig. 7}(b)), as more sites are available to retain O\textsubscript{2} molecules. However, the exact desorption mechanism appears complex and may involve both O\textsubscript{2} desorption resulting from the reorganization of the water ice matrix (e.g., volcano desorption upon crystallization) and co-desorption of O\textsubscript{2} molecules with water. For completeness, Appendix~\ref{app:C} shows the results of the control experiments using isotopically labelled dioxygen ices, which agree with the experiments performed for the main isotopologue.

\begin{figure}
  \resizebox{\hsize}{!}{\includegraphics{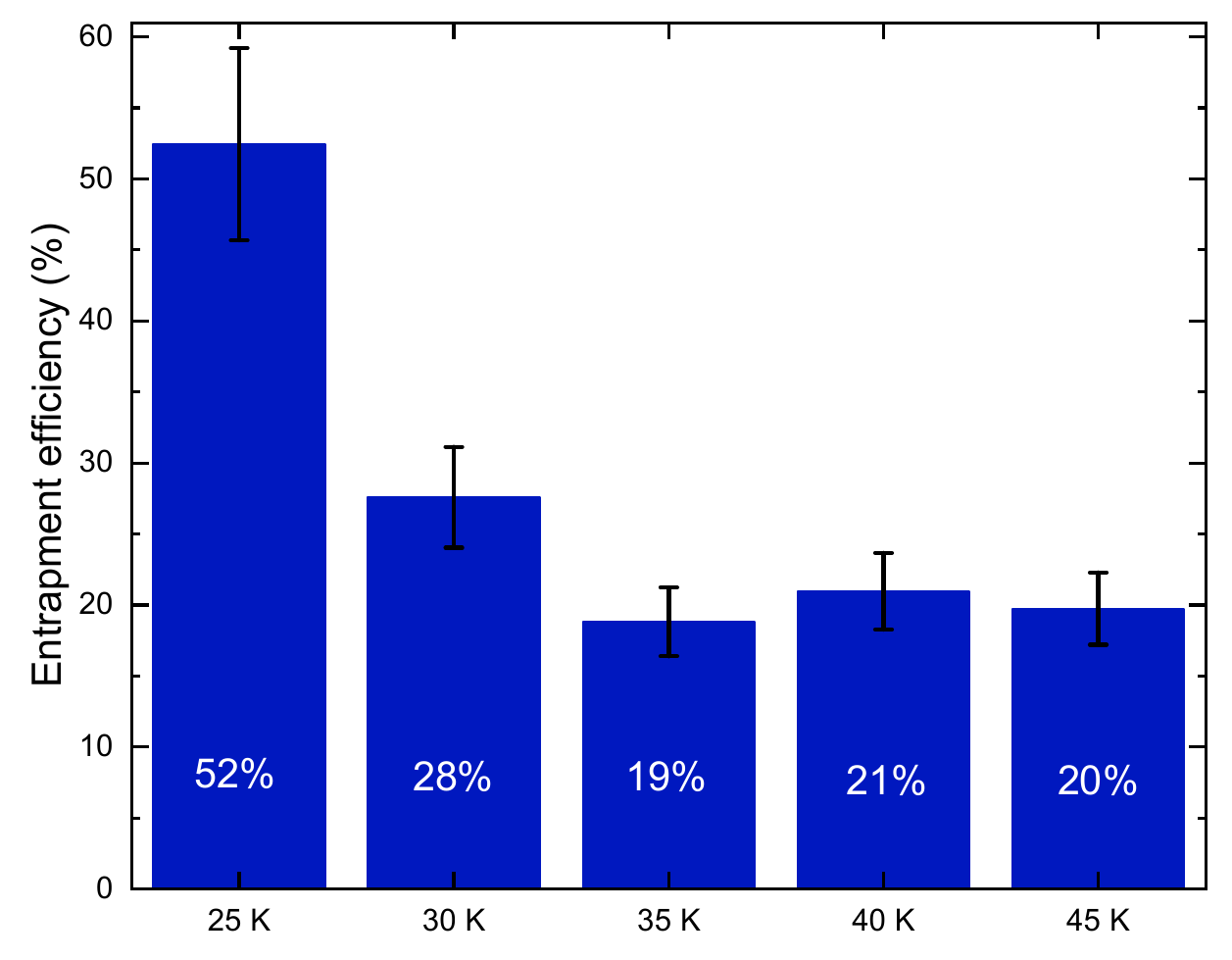}}
  \caption{Corrected entrapment efficiencies of O\textsubscript{2} in ASW ice derived from experiments performed at different isothermal temperatures (25–45 K). Error bars were estimated as $\pm$13\% of measured values.}
  \label{Fig. 8}
\end{figure}

The derived entrapment efficiencies of O\textsubscript{2} in ASW from experiments 1 to 5, determined as described in Sect.~\ref{sec:2.3.2}, are plotted in Fig.~\ref{Fig. 8}, with their 13\% relative uncertainty shown as error bars. The efficiencies decrease exponentially, from an initial value of (52 $\pm$ 7)\% to a lower-limit of about 20\%, a value that remains constant for all temperatures above 35 K. A similar entrapment efficiency of (23 $\pm$ 3)\% is obtained for experiment 4e, conducted with a 30-hour isothermal phase at 40 K and a 40 ML water ice layer. Therefore, even with long diffusion times or elevated temperatures, a relatively constant fraction of about 20\% of O\textsubscript{2} molecules can remain trapped in ASW.  

\section{Astrophysical implications}

\label{sec:4}

Studying the surface diffusion of O\textsubscript{2} in porous ASW ice analogues at low temperatures ($\leq$45 K) helps to better understand interstellar ice dynamics in dense molecular clouds. In astrochemical models, the diffusion energy barrier is a highly sensitive parameter, and its accurate determination significantly improves the reliability of simulations aimed at studying the ice chemistry of prestellar objects, particularly during the protostar stages \citep{dijkhuis2025, ligterink2025}. Because of the lack of experimental measurements of diffusion coefficients and, consequently, diffusion energy barriers, astrochemical models often rely on generic ratios $\chi$ = $\frac{E_{\mathrm{D}}}{E_{\mathrm{des}}}$ \citep{tielens1982, garrod2011, chang2012} without explicitly accounting for the specific species, the ice morphology, or the physical conditions. For example, \citet{minissale2013} considered $\chi$ = 0.7, while \citet{taquet2016} adopted an average value of 0.5, giving very different results on O\textsubscript{2} dynamics. 

In this study, we derived O\textsubscript{2} diffusion coefficients on the order of 10\textsuperscript{$-$16} to 10\textsuperscript{$-$15} cm\textsuperscript{2} s\textsuperscript{$-$1} for temperatures between 35 K and 45 K (Table~\ref{table:3}). These values are consistent with those reported in the literature for other small hypervolatiles such as CO, which span 10\textsuperscript{$-$18} to 10\textsuperscript{$-$12} cm\textsuperscript{2} s\textsuperscript{$-$1} over the temperature range 12–50 K \citep{karssemeijer2013, lauck2015}. Using the Arrhenius law, we extracted a diffusion energy barrier of 10 $\pm$ 3 meV (116 $\pm$ 35 K) for O\textsubscript{2} in ASW, which is lower than the experimental value reported by \citet{he2018}, who measured $E_{\mathrm{D}}$ =  38 $\pm$ 1 meV (446 $\pm$ 12 K), by monitoring the effect of O\textsubscript{2} diffusion on the dangling-OH band strength of  H\textsubscript{2}O ice. This work provides a direct laboratory constraint on the diffusion energy of IR-inactive O\textsubscript{2} molecules by applying Fick’s law to its quantified diffusion-limited desorption, an approach that is less subject to additional uncertainties introduced by extra-layers of inferences. Regardless, the determination of the diffusion energy barrier can yield distributed results, as observed in studies on CO diffusion in ASW, with values ranging from 10 $\pm$ 15 to 30 $\pm$ 4 meV (116 $\pm$ 174 to 350 $\pm$ 50 K) \citep{mispelaer2013, karssemeijer2013, lauck2015, kouchi2020}. Such variability arises because even slight differences in experimental conditions can alter the ice structure and, consequently, the measured diffusion energy barrier. The relatively low value for O\textsubscript{2} diffusion energy barrier reveals that this process is easily activated, even at low temperatures, a behavior also observed for H atoms ($E_{\mathrm{D}}$ = 20 meV, \citep{watanabe2010}) or CO molecules ($E_{\mathrm{D}}$ = 14 $\pm$ 1 meV, \citep{lauck2015}). Based on $E_{\mathrm{des}}$ = 1107 K, given by \citet{minissale2022} for the desorption energy barrier of O\textsubscript{2}, we experimentally determined $\chi$ to lie between 0.07 and 0.14. This ratio is clearly smaller than the values typically used in astrochemical models for O\textsubscript{2} dynamics, highlighting the importance of using specific experimentally determined ratios. Nevertheless, note that no desorption energy was determined in this study, which introduces uncertainties stemming from potential differences in experimental conditions that could affect this value. Moreover, the $\chi$ ratios strongly depend on the solid-state conditions and the molecule. For instance, \citet{furuya2022} experimentally found $\chi$ to range between 0.14 and 0.73 for CH\textsubscript{4}, H\textsubscript{2}S, OCS, CH\textsubscript{3}CN, and CH\textsubscript{3}OH, showing ratios can hardly be predicted unless experimentally derived under the relevant conditions.  

We determined a minimal entrapment fraction of 20\% for O\textsubscript{2} in ASW at the highest temperatures and longest timescales of the isothermal phase. Because the entrapment efficiency reaches a plateau, we demonstrated that, even in the ISM where heating timescales are much longer than in the laboratory, interstellar ices may always contain a residual amount of entrapped volatiles such as O\textsubscript{2}, which will remain stuck in the ice until desorbing with water. This experimental finding is consistent with previous observations e.g. highlighting the importance of taking into account entrapment to understand hypervolatiles distribution across protoplanetary disks \citep{bergner2024}. 

\section{Conclusions}

\label{sec:5}

In this study, the surface diffusion behavior of O\textsubscript{2} in porous ASW has been quantified by monitoring, for the first time, the diffusion-limited desorption of O\textsubscript{2} molecules using mass spectrometry. This new approach enables a comprehensive characterization of the diffusion of IR-inactive molecules in a more straightforward way, while remaining agnostic of the ice matrix, i.e. it does not rely on the IR monitoring of another deposited species. The surface diffusion energy barrier was determined by extracting the diffusion coefficients at different isothermal temperatures (35 K, 40 K and 45 K), and the O\textsubscript{2} entrapment efficiencies in the ASW were also measured. The main experimental findings are concluded as follows:
\begin{enumerate}
    \item We obtain diffusion coefficients for O\textsubscript{2} molecules in porous ASW at different astronomically relevant temperatures. From 35 to 45 K, the diffusion coefficients range from (0.5 $\pm$ 0.2) to (1.0 $\pm$ 0.3) $\times$ 10\textsuperscript{$-$15} cm\textsuperscript{2} s\textsuperscript{$-$1}. With increasing water coverage, O\textsubscript{2} diffusion coefficients range between (0.7$-$2) $\times$ 10\textsuperscript{$-$15} cm\textsuperscript{2} s\textsuperscript{$-$1}. 
    
\item By fitting the Arrhenius equation, we derive a diffusion energy barrier of 10 $\pm$ 3 meV (116 $\pm$ 35 K) for the diffusion of O\textsubscript{2} in ASW, with a pre-exponential factor of $D_{0}$ = (2.3 $\pm$ 1.9) $\times$ 10\textsuperscript{$-$14} cm\textsuperscript{2} s\textsuperscript{$-$1}. Compared to other small volatiles, this energy barrier is relatively low, yielding a $\chi$ ratio of $\sim$0.1, which highlights an easier mobility of O\textsubscript{2} molecules in the water matrix than previously expected. 

\item The entrapment efficiency of O\textsubscript{2} in ASW ice shows a strong temperature dependence from 25 K to 35 K, decreasing from $\sim$52\% to $\sim$20\%. However, the entrapment fraction remains roughly constant at $\sim$20\% in laboratory measurements up to 45 K, or even over relatively long timescales.

\end{enumerate}

These results provide new constraints on O\textsubscript{2} surface diffusion and entrapment in porous ASW. The characterised diffusion coefficients and energy barrier will help in astrochemical modeling, which emphasizes the importance of accurately determining them for future model implementations. It also opens possibilities for future studies aimed at quantifying the diffusion parameters of other IR-inactive molecules such as N\textsubscript{2} or noble gases. On another hand, the entrapment efficiencies can be further investigated using higher isothermal temperatures and other hypervolatiles to confirm the trend, in which small molecules, such as O\textsubscript{2}, remain trapped within the ice matrix well above their nominal sublimation temperature.

\begin{acknowledgements}
L. C. was supported by the ERASMUS+ grant financed by the European Union.
    J. C. S. was supported by the Heising-Simons Foundation through a 51 Pegasi b Fellowship and by the Danish National Research Foundation through the Center of Excellence “InterCat” (Grant agreement no.: DNRF150)
\end{acknowledgements}

%
\bibliographystyle{aa} 
\bibliography{reference}

@article{acharyya2022,
  title={Understanding the impact of diffusion of CO in the astrochemical models},
  author={Acharyya, Kinsuk},
  year={2022},
  publisher={Cambridge University Press},
note= {\href{https://ui.adsabs.harvard.edu/abs/2022PASA...39....9A/abstract}{PASA, 39, e009}}
}

@article{bergner2024,
  title={JWST ice band profiles reveal mixed ice compositions in the HH 48 NE disk},
  author={Bergner, Jennifer B and Sturm, JA and Piacentino, Elettra L and McClure, MK and {\"O}berg, Karin I and Boogert, ACA and Dartois, E and Drozdovskaya, MN and Fraser, HJ and Harsono, Daniel and others},
  year={2024},
  publisher={IOP Publishing},
note={\href{https://ui.adsabs.harvard.edu/abs/2024ApJ...975..166B/abstract}{ApJ, 975, 166}}
}

@article{boogert2015,
  title={Observations of the icy universe},
  author={Boogert, AC Adwin and Gerakines, Perry A and Whittet, Douglas CB},
  journal={Annual Review of Astronomy and Astrophysics},
  year={2023},
  publisher={Annual Reviews},
note={\href{https://ui.adsabs.harvard.edu/abs/2015ARA%26A..53..541B/abstract}{Annu. Rev. Astron. Astrophys., 53, 541}}
}

@article{bossa2012,
  title={Thermal collapse of porous interstellar ice},
  author={Bossa, J-B and Isokoski, K and De Valois, MS and Linnartz, H},
  year={2012},
  publisher={EDP Sciences},
note={\href{https://ui.adsabs.harvard.edu/abs/2012A&A...545A..82B}{A\&A, 545, A82}}
}

@article{brown1996,
  title={H2O condensation coefficient and refractive index for vapor-deposited ice from molecular beam and optical interference measurements},
  author={Brown, DE and George, Steven M and Huang, Chen and Wong, EKL and Rider, Keith B and Smith, R Scott and Kay, Bruce D},
  year={1996},
  publisher={ACS Publications},
note={\href{https://ui.adsabs.harvard.edu/abs/1996JPhCh.100.4988B/abstract}{J. Phys. Chem., 100, 4988}}
}

@article{chang2012,
  title={A unified microscopic--macroscopic monte carlo simulation of gas-grain chemistry in cold dense interstellar clouds},
  author={Chang, Qiang and Herbst, Eric},
  year={2012},
  publisher={IOP Publishing},
note={\href{https://ui.adsabs.harvard.edu/abs/2012ApJ...759..147C/abstract}{ApJ, 759, 147}}
}

@phdthesis{chuang2018,
  title={The formation of complex organic molecules in dense clouds: sweet results from the laboratory},
  author={Chuang, K.-J.},
  year={2018},
note= {\href{https://scholarlypublications.universiteitleiden.nl/access/item%3A2904345/view}{Leiden University}}
}

@article{cuppen2010,
  title={Water formation at low temperatures by surface O 2 hydrogenation II: the reaction network},
  author={Cuppen, HM and Ioppolo, S and Romanzin, C and Linnartz, H},
  year={2010},
  publisher={Royal Society of Chemistry},
note={\href{https://ui.adsabs.harvard.edu/abs/2010PCCP...1212077C/abstract}{Phys. Chem. Chem. Phys., 12, 12077}}
}

@article{cuppen2017,
  title={Grain surface models and data for astrochemistry},
  author={Cuppen, HM and Walsh, C and Lamberts, T and Semenov, D and Garrod, RT and Penteado, E Monfardini and Ioppolo, S},
  year={2017},
  publisher={Springer},
note={\href{https://ui.adsabs.harvard.edu/abs/2017SSRv..212....1C/abstract}{Space Sci. Rev., 212, 1}}
}

@article{cuppen2024,
  title={Laboratory and computational studies of interstellar ices},
  author={Cuppen, Herma M and Linnartz, Harold and Ioppolo, Sergio},
  year={2024},
  publisher={Annual Reviews},
note={\href{https://ui.adsabs.harvard.edu/abs/2024ARA%26A..62..243C/abstract}{Annu. Rev. Astron. Astrophys., 62}}
}

@article{dartois2005,
  title={The ice survey opportunity of ISO},
  author={Dartois, Emmanuel},
  year={2005},
  publisher={Springer},
note={\href{https://ui.adsabs.harvard.edu/abs/2005SSRv..119..293D/abstract}{Space Sci. Rev., 119, 293}}
}

@article{dijkhuis2025,
  title={A sensitivity analysis of interstellar ice chemistry in astrochemical models},
  author={Dijkhuis, Tobias M and Lamberts, Thanja and Viti, Serena and Cuppen, Herma M},
  year={2025},
note={\href{https://arxiv.org/pdf/2511.01042}{A\&A, Submitted}}
}

@article{fuchs2009,
  title={Hydrogenation reactions in interstellar CO ice analogues-A combined experimental/theoretical approach},
  author={Fuchs, GW and Cuppen, HM and Ioppolo, S and Romanzin, C and Bisschop, SE and Andersson, S and Van Dishoeck, EF and Linnartz, H},
  year={2009},
  publisher={EDP Sciences},
note={\href{https://ui.adsabs.harvard.edu/abs/2009arXiv0906.2292F/abstract}{A\&A, 505, 629}}
}

@article{fulvio2009,
  title={Novel measurements of refractive index, density and mid-infrared integrated band strengths for solid O2, N2O and NO2: N2O4 mixtures},
  author={Fulvio, D and Sivaraman, B and Baratta, GA and Palumbo, ME and Mason, NJ},
  year={2009},
  publisher={Elsevier},
note = {\href{https://ui.adsabs.harvard.edu/abs/2009AcSpA..72.1007F/abstract}{Spectrochim. Acta, Part A, 72, 1007}}
}

@article{furuya2022,
  title={Diffusion activation energy and desorption activation energy for astrochemically relevant species on water ice show no clear relation},
  author={Furuya, Kenji and Hama, Tetsuya and Oba, Yasuhiro and Kouchi, Akira and Watanabe, Naoki and Aikawa, Yuri},
  year={2022},
  publisher={IOP Publishing},
note={\href{https://ui.adsabs.harvard.edu/abs/2022ApJ...933L..16F/abstract}{ApJL, 933, L16}}
}

@article{garrod2011,
  title={On the formation of CO2 and other interstellar ices},
  author={Garrod, Robin T and Pauly, Tyler},
  year={2011},
  publisher={IOP Publishing},
note={\href{http://ui.adsabs.harvard.edu/abs/2011ApJ...735...15G/abstract}{ApJ, 735, 15}}
}

@article{gent2013,
  title={The supernova-regulated ISM--I. The multiphase structure},
  author={Gent, FA and Shukurov, A and Fletcher, A and Sarson, GR and Mantere, MJ},
  year={2013},
  publisher={The Royal Astronomical Society},
note={\href{https://ui.adsabs.harvard.edu/abs/2013MNRAS.432.1396G/abstract}{Mon. Not. R. Astron. Soc., 432, 1396}}
}

@article{gerakines1995,
  title={The infrared band strengths of H2O, CO and CO2 in laboratory simulations of astrophysical ice mixtures},
  author={Gerakines, Perry A and Schutte, WA and Greenberg, JM and van Dishoeck, Ewine F},
  year={1995},
note={\href{https://ui.adsabs.harvard.edu/abs/1995A%26A...296..810G/abstract}{A\&A, 296, A810}}
}

@article{hasegawa1992,
  title={Models of gas-grain chemistry in dense interstellar clouds with complex organic molecules},
  author={Hasegawa, Tatsuhiko I and Herbst, Eric and Leung, Chun M},
  year={1992},
note={\href{https://adsabs.harvard.edu/full/1992ApJS...82..167H}{ApJS, 82, 167}}
}

@article{he2018,
  title={Measurements of diffusion of volatiles in amorphous solid water: application to interstellar medium environments},
  author={He, Jiao and Emtiaz, SM and Vidali, Gianfranco},
  year={2018},
  publisher={IOP Publishing},
note={\href{https://ui.adsabs.harvard.edu/abs/2018ApJ...863..156H/abstract}{ApJ, 863, 156}}
}

@article{herbst1973,
  title={The formation and depletion of molecules in dense interstellar clouds},
  author={Herbst, Eric and Klemperer, William},
  year={1973},
note={\href{https://adsabs.harvard.edu/full/1973ApJ...185..505H}{ApJ, 185, 505}}
}

@article{inostroza2024,
  title={Formation pathways of formic acid (HCOOH) in regions with methanol ices},
  author={Inostroza-Pino, Natalia and Godwin, Oko Emmanuel and Mardones, Diego and Ge, Jixing},
  year={2024},
  publisher={EDP Sciences},
note={\href{https://ui.adsabs.harvard.edu/abs/2024A%26A...688A.140I/abstract}{A\&A, 688, A140}}
}

@article{ioppolo2010,
  title={Water formation at low temperatures by surface O 2 hydrogenation I: characterization of ice penetration},
  author={Ioppolo, Sergio and Cuppen, Herma M and Romanzin, Claire and van Dishoeck, Ewine F and Linnartz, Harold},
  year={2010},
  publisher={Royal Society of Chemistry},
note={\href{https://ui.adsabs.harvard.edu/abs/2010PCCP...1212065I/abstract}{Phys. Chem. Chem. Phys., 12, 12065}}
}

@article{ioppolo2013,
  title={SURFRESIDE2: An ultrahigh vacuum system for the investigation of surface reaction routes of interstellar interest},
  author={Ioppolo, S and Fedoseev, G and Lamberts, T and Romanzin, C and Linnartz, H},
  year={2013},
  publisher={AIP Publishing},
note={\href{https://ui.adsabs.harvard.edu/abs/2013RScI...84g3112I/abstract}{Rev. Sci. Instrum., 84}}
}

@article{itikawa2005,
  title={Cross sections for electron collisions with water molecules},
  author={Itikawa, Yukikazu and Mason, Nigel},
  year={2005},
  publisher={American Institute of Physics for the National Institute of Standards and~…},
note={\href{https://ui.adsabs.harvard.edu/abs/2005JPCRD..34....1I/abstract}{J. Phys. Chem. Ref. Data, 34, 1}}
}

@article{karssemeijer2013,
  title={Dynamics of CO in amorphous water-ice environments},
  author={Karssemeijer, LJ and Ioppolo, S and Van Hemert, MC and Van Der Avoird, A and Allodi, MA and Blake, GA and Cuppen, HM},
  year={2013},
  publisher={IOP Publishing},
note={\href{https://ui.adsabs.harvard.edu/abs/2014ApJ...781...16K/abstract}{ApJ, 781, 16}}
}

@article{kouchi2020,
  title={Direct measurements of activation energies for surface diffusion of CO and CO2 on amorphous solid water using in situ transmission electron microscopy},
  author={Kouchi, Akira and Furuya, Kenji and Hama, Tetsuya and Chigai, Takeshi and Kozasa, Takashi and Watanabe, Naoki},
  year={2020},
  publisher={IOP Publishing},
note={\href{https://ui.adsabs.harvard.edu/abs/2020ApJ...891L..22K/abstract}{ApJL, 891, L22}}
}

@article{krishnakumar1992,
  title={Cross-sections for electron impact ionization of O2},
  author={Krishnakumar, E and Srivastava, SK},
  year={1992},
  publisher={Elsevier},
note={\href{https://ui.adsabs.harvard.edu/abs/1992IJMSI.113....1K/abstract}{Int. J. Mass Spectrom. Ion Process., 113, 1}}
}

@article{lammer2009,
  title={What makes a planet habitable?},
  author={Lammer, Helmut and Bredeh{\"o}ft, JH and Coustenis, A and Khodachenko, ML and Kaltenegger, L and Grasset, O and Prieur, D and Raulin, F and Ehrenfreund, P and Yamauchi, M and others},
  year={2009},
  publisher={Springer},
note={\href{https://ui.adsabs.harvard.edu/abs/2009A%26ARv..17..181L/abstract}{Astron. Astrophys. Rev., 17, 181}}
}

@article{lauck2015,
  title={CO diffusion into amorphous H2O ices},
  author={Lauck, Trish and Karssemeijer, Leendertjan and Shulenberger, Katherine and Rajappan, Mahesh and {\"O}berg, Karin I and Cuppen, Herma M},
  year={2015},
  publisher={IOP Publishing},
note={\href{https://ui.adsabs.harvard.edu/abs/2015ApJ...801..118L/abstract}{ApJ, 801, 118}}
}

@article{ligterink2025,
  title={Molecular mobility of extraterrestrial ices: surface diffusion in astrochemistry and planetary science},
  author={Ligterink, NFW and Walsh, C and Cuppen, HM and Drozdovskaya, MN and Ahmad, A and Benoit, DM and Carder, JT and Das, A and D{\'\i}az-Berr{\'\i}os, JK and Dulieu, F and others},
  year={2025},
  publisher={Royal Society of Chemistry},
note={\href{https://ui.adsabs.harvard.edu/abs/2025PCCP...2719630L/abstract}{Phys. Chem. Chem. Phys., 27, 19630}}
}

@article{linnartz2015,
  title={Atom addition reactions in interstellar ice analogues},
  author={Linnartz, Harold and Ioppolo, Sergio and Fedoseev, Gleb},
  year={2015},
  publisher={Taylor \& Francis},
note={\href{https://ui.adsabs.harvard.edu/abs/2015arXiv150702729L/abstract}{Int. Rev. Phys. Chem., 34, 205}}
}

@article{martin2015,
  title={UV photoprocessing of CO2 ice: a complete quantification of photochemistry and photon-induced desorption processes},
  author={Mart{\'\i}n-Dom{\'e}nech, R and Manzano-Santamar{\'\i}a, J and Caro, GM Mu{\~n}oz and Cruz-D{\'\i}az, Gustavo A and Chen, Y-J and Herrero, V{\'\i}ctor J and Tanarro, Isabel},
  year={2015},
  publisher={EDP Sciences},
note={\href{https://ui.adsabs.harvard.edu/abs/2015A%26A...584A..14M/abstract}{A\&A, 584, A14}}
}

@article{mate2012,
  title={Morphology and crystallization kinetics of compact (HGW) and porous (ASW) amorphous water ice},
  author={Mat{\'e}, Bel{\'e}n and Rodr{\'\i}guez-Lazcano, Yamilet and Herrero, Victor J},
  year={2012},
  publisher={Royal Society of Chemistry},
note={\href{https://ui.adsabs.harvard.edu/abs/2012PCCP...1410595M/abstract}{Phys. Chem. Chem. Phys., 14, 10595}}
}

@article{mate2020,
  title={Diffusion of CH4 in amorphous solid water},
  author={Mat{\'e}, Bel{\'e}n and Cazaux, St{\'e}phanie and Satorre, Miguel {\'A}ngel and Molpeceres, Germ{\'a}n and Ortigoso, Juan and Mill{\'a}n, Carlos and Santonja, Carmina},
  year={2020},
  publisher={EDP Sciences},
note={\href{https://ui.adsabs.harvard.edu/abs/2020%A26A...643A.163M/abstract}{A\&A, 643, A163}}
}

@article{mcclure2023,
  title={An Ice Age JWST inventory of dense molecular cloud ices},
  author={McClure, Melissa K and Rocha, WRM and Pontoppidan, KM and Crouzet, N and Chu, Laurie EU and Dartois, E and Lamberts, T and Noble, JA and Pendleton, YJ and Perotti, G and others},
  year={2023},
  publisher={Nature Publishing Group UK London},
note={\href{https://ui.adsabs.harvard.edu/abs/2023NatAs...7..431M/abstract}{Nat. Astron., 7, 431}}
}

@article{minissale2013,
  title={Quantum tunneling of oxygen atoms on very cold surfaces},
  author={Minissale, M and Congiu, E and Baouche, S and Chaabouni, H and Moudens, A and Dulieu, Francois and Accolla, Mario and Cazaux, S and Manic{\'o}, G and Pirronello, Valerio},
  year={2013},
  publisher={APS},
note={\href{https://ui.adsabs.harvard.edu/abs/2013PhRvL.111e3201M/abstract}{Phys. Rev. Lett., 111, 053201}}
}

@article{minissale2016a,
  title={Dust as interstellar catalyst-I. Quantifying the chemical desorption process},
  author={Minissale, M and Dulieu, Francois and Cazaux, S and Hocuk, S},
  year={2016},
  publisher={EDP Sciences},
note={\href{https://ui.adsabs.harvard.edu/abs/2016A%26A...585A..24M/abstract}{A\&A, 585, A24}}
}

@article{minissale2016b,
  title={Direct measurement of desorption and diffusion energies of O and N atoms physisorbed on amorphous surfaces},
  author={Minissale, Marco and Congiu, Emanuele and Dulieu, Fran{\c{c}}ois},
  year={2016},
  publisher={EDP Sciences},
note={\href{https://ui.adsabs.harvard.edu/abs/2016A%26A...585A.146M/abstract}{A\&A, 585, A146}}
}

@article{minissale2022,
  title={Thermal desorption of interstellar ices: A review on the controlling parameters and their implications from snowlines to chemical complexity},
  author={Minissale, Marco and Aikawa, Yuri and Bergin, Edwin and Bertin, Mathieu and Brown, Wendy A and Cazaux, Stephanie and Charnley, Steven B and Coutens, Audrey and Cuppen, Herma M and Guzman, Victoria and others},
  year={2022},
  publisher={ACS Publications},
note={\href{https://ui.adsabs.harvard.edu/abs/2022ESC.....6..597M/abstract}{ACS Earth Space Chem., 6, 597}}
}

@article{mispelaer2013,
  title={Diffusion measurements of CO, HNCO, H2CO, and NH3 in amorphous water ice},
  author={Mispelaer, F and Theul{\'e}, P and Aouididi, H and Noble, J and Duvernay, F and Danger, Gregoire and Roubin, P and Morata, O and Hasegawa, T and Chiavassa, T},
  year={2013},
  publisher={EDP Sciences},
note={\href{https://ui.adsabs.harvard.edu/abs/2013A%26A...555A..13M/abstract}{A\&A, 555, A13}}
}

@article{oberg2009,
  title={Quantification of segregation dynamics in ice mixtures},
  author={{\"O}berg, Karin I and Fayolle, Edith C and Cuppen, Herma M and van Dishoeck, Ewine F and Linnartz, Harold},
  year={2009},
  publisher={EDP Sciences},
note={\href{https://ui.adsabs.harvard.edu/abs/2009A%26A...505..183O/abstract}{A\&A, 505, 183}}
}

@article{oberg2021,
  title={Astrochemistry and compositions of planetary systems},
  author={{\"O}berg, Karin I and Bergin, Edwin A},
  year={2021},
  publisher={Elsevier},
note={\href{https://ui.adsabs.harvard.edu/abs/2021PhR...893....1O/abstract}{Phys. Rep., 893, 1}}
}

@phdthesis{santosthesis,
    author = {Santos, Julia C},
    title = {Transformation and sublimation of interstellar ices},
    year = {2025},
note ={\href{https://scholarlypublications.universiteitleiden.nl/access/item%3A4252327/view}{Leiden University}}
}

@article{santos2023,
  title={Interaction of H2S with H atoms on grain surfaces under molecular cloud conditions},
  author={Santos, Julia C and Linnartz, Harold and Chuang, K-J},
  year={2023},
  publisher={EDP Sciences},
note={\href{https://ui.adsabs.harvard.edu/abs/2023A%26A...678A.112S/abstract}{A\&A, 678, A112}}
}

@article{slavicinska2024,
  title={JWST detections of amorphous and crystalline HDO ice toward massive protostars},
  author={Slavicinska, Katerina and Van Dishoeck, Ewine F and Tychoniec, {\L}ukasz and Nazari, Pooneh and Rubinstein, Adam E and Gutermuth, Robert and Tyagi, Himanshu and Chen, Yuan and Brunken, Nashanty GC and Rocha, Will RM and others},
  year = {2024},
  publisher={EDP Sciences},
note={\href{https://ui.adsabs.harvard.edu/abs/2024A%26A...688A..29S/abstract}{A\&A, 688, A29}}
}

@article{taquet2016,
  title={A primordial origin for molecular oxygen in comets: a chemical kinetics study of the formation and survival of O2 ice from clouds to discs},
  author={Taquet, Vianney and Furuya, Kenji and Walsh, Catherine and van Dishoeck, Ewine F},
  year={2016},
  publisher={The Royal Astronomical Society},
note={\href{https://ui.adsabs.harvard.edu/abs/2016MNRAS.462S..99T/abstract}{Mon. Not. R. Astron. Soc., 462, S99}}
}

@article{tielens1982,
  title={Model calculations of the molecular composition of interstellar grain mantles},
  author={Tielens, AGGM and Hagen, W},
  year={1982},
note={\href{https://ui.adsabs.harvard.edu/abs/1982A%26A...114..245T/abstract}{A\&A, 114, 245}}
}

@article{rimola2014,
  title={Combined quantum chemical and modeling study of CO hydrogenation on water ice},
  author={Rimola, Albert and Taquet, Vianney and Ugliengo, Piero and Balucani, Nadia and Ceccarelli, Cecilia},
  year={2014},
  publisher={EDP Sciences},
note={\href{https://ui.adsabs.harvard.edu/abs/2014A%26A...572A..70R/abstract}{A\&A, 572, A70}}
}

@article{van2014,
  title={Astrochemistry of dust, ice and gas: introduction and overview},
  author={van Dishoeck, Ewine F},
  year={2014},
  publisher={Royal Society of Chemistry},
note={\href{https://ui.adsabs.harvard.edu/abs/2014FaDi..168....9V/abstract}{Faraday Discuss., 168, 9}}
}

@article{watanabe2010,
  title={Direct measurements of hydrogen atom diffusion and the spin temperature of nascent H2 molecule on amorphous solid water},
  author={Watanabe, Naoki and Kimura, Yuki and Kouchi, Akira and Chigai, Takeshi and Hama, Tetsuya and Pirronello, Valerio},
  year={2010},
  publisher={IOP Publishing},
note={\href{https://ui.adsabs.harvard.edu/abs/2010ApJ...714L.233W/abstract}{ApJL, 714, L233}}
}







   
  



\begin{appendix}




\twocolumn
\section{$k_{\mathrm{QMS}}$ determination}
\label{app:A}
The proportionality constant $k_{\mathrm{QMS}}$ for this setup was determined by comparing the IR-derived H\textsubscript{2}O column densities ($N_{\mathrm{H_{2}O}}$) with the $\frac{N_{\mathrm{H_{2}O}}}{k_{\mathrm{QMS}}}$ values from QMS data across all experiments. A mean $k_{\mathrm{QMS}}$ value was then extracted along with its standard deviation, yielding $k_{\mathrm{QMS}}$  = (7.89 $\pm$ 1.76) $\times$ $10^{21}$ a.u.
%


\section{Atmospheric O\textsubscript{2} contamination in water sample}
\begin{figure}[h!]
  \resizebox{\hsize}{!}{\includegraphics{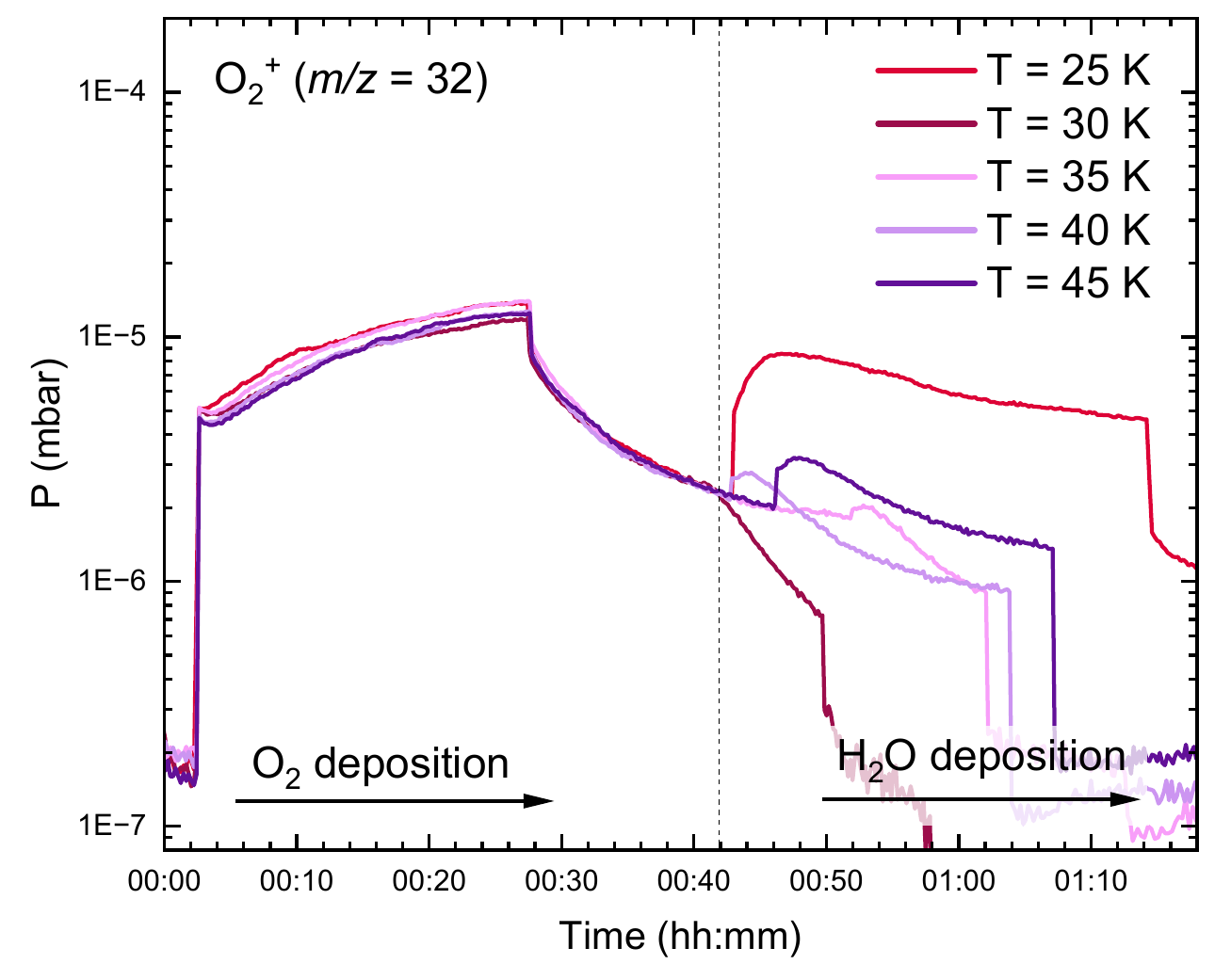}}
  \caption{QMS signals at $m/z$ = 32 recorded during the consecutive deposition of O\textsubscript{2} and H\textsubscript{2}O for five experiments with different isothermal temperatures (T = 25 K, 30 K, 35 K, 40 K, 45 K). The presence of a signal at $m/z$ = 32 during H\textsubscript{2}O deposition indicates contamination of the water sample by atmospheric O\textsubscript{2}.}
  \label{Fig. B.1}
\end{figure}

\begin{figure}[h!]
  \resizebox{\hsize}{!}{\includegraphics{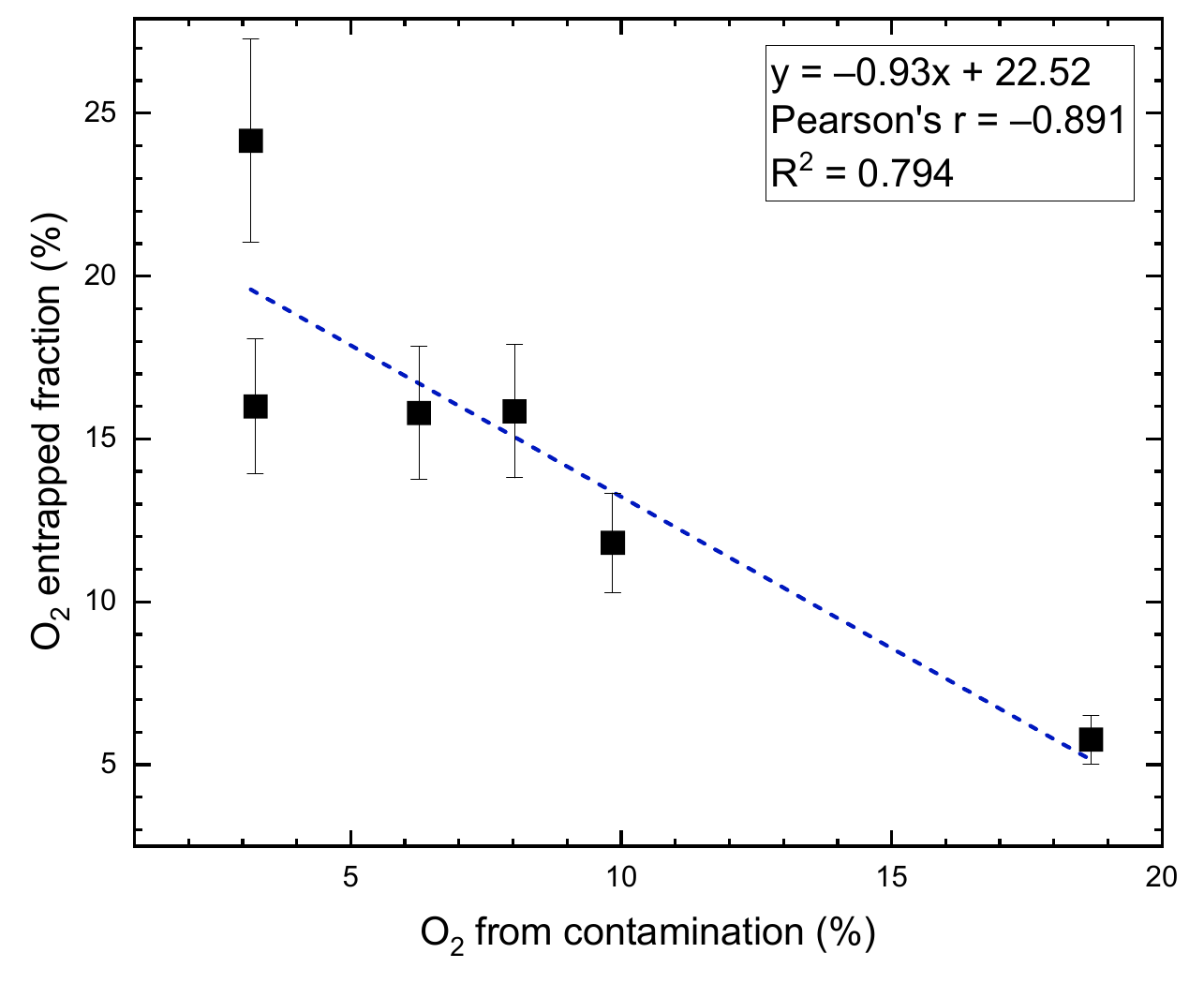}}
  \caption{Linear correlation between fraction of entrapped O\textsubscript{2} and the fraction coming from contamination by atmospheric air. The fit is used to correct entrapment efficiencies by extrapolation at 0\% contamination.}
  \label{Fig. B.2}
\end{figure}

Contamination of the water samples by atmospheric O\textsubscript{2} is further quantified in this Appendix. Fig.~\ref{Fig. B.1} illustrates how the level of contamination varies across experiments, while Fig.~\ref{Fig. B.2} shows the linear correlation between contamination and the final amount of entrapped O\textsubscript{2}. This correlation allows the derivation of a correction factor, subsequently applied to all entrapment efficiency values.
\FloatBarrier

\FloatBarrier 

\section{Isotopically labeled experiments}
\label{app:C}
\begin{figure}[h!]
  \resizebox{\hsize}{!}{\includegraphics{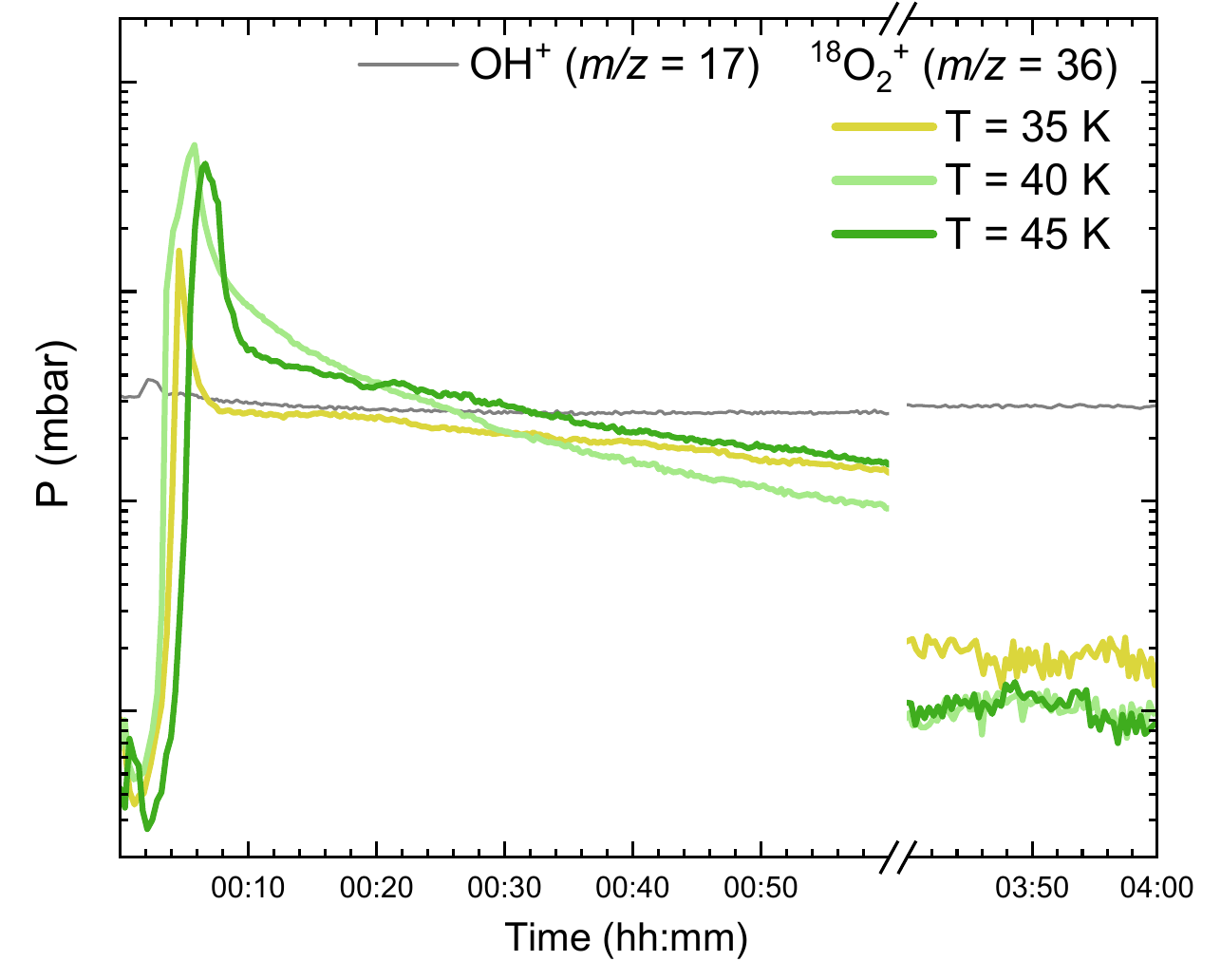}}
  \caption{QMS signals recorded during the four-hour isothermal phase at $m/z$ = 17 (OH\textsuperscript{+}) and $m/z$ = 36 (\textsuperscript{18}O\textsubscript{2}\textsuperscript{+}) for experiments 8 to 10 at different isothermal temperatures (35 K to 45 K with 40 ML of H\textsubscript{2}O). Considering the signal at $m/z$ = 17 was the same for every experiment, only the water profile of experiment 9 (T = 40 K) is shown for clarity.}
  \label{Fig. C.1}
\end{figure}

Fig.~\ref{Fig. C.1} shows the QMS signal at $m/z$ = 36 (\textsuperscript{18}O\textsubscript{2}\textsuperscript{+}) and $m/z$ = 17 (OH\textsuperscript{+}) during the isothermal phase of the isotopic experiments. Similar to what was observed in the non-isotopic experiments, the signal corresponding to the isotopic molecular oxygen cation exhibits a sharp peak at the beginning of the isothermal phase, followed by a similar exponential decay, while water displays a constant profile. This analogous behavior indicates that the diffusion phenomenon is independent of the isotopologue and that possible contamination does not affect the determination of the diffusion coefficients, i.e., confirming that the observed signal for the major isotopologue arises from diffusion through the water ice matrix followed by instantaneous desorption upon reaching the surface. However, in the isotopic experiments, the initial desorption peak does not show a clear evolution, but as \textsuperscript{18}O\textsubscript{2}/H\textsubscript{2}O ratios were systematically lower than the O\textsubscript{2}/H\textsubscript{2}O ones, it may cause slight differences in behavior. 

\begin{figure*}[h!]
  \resizebox{\hsize}{!}{\includegraphics{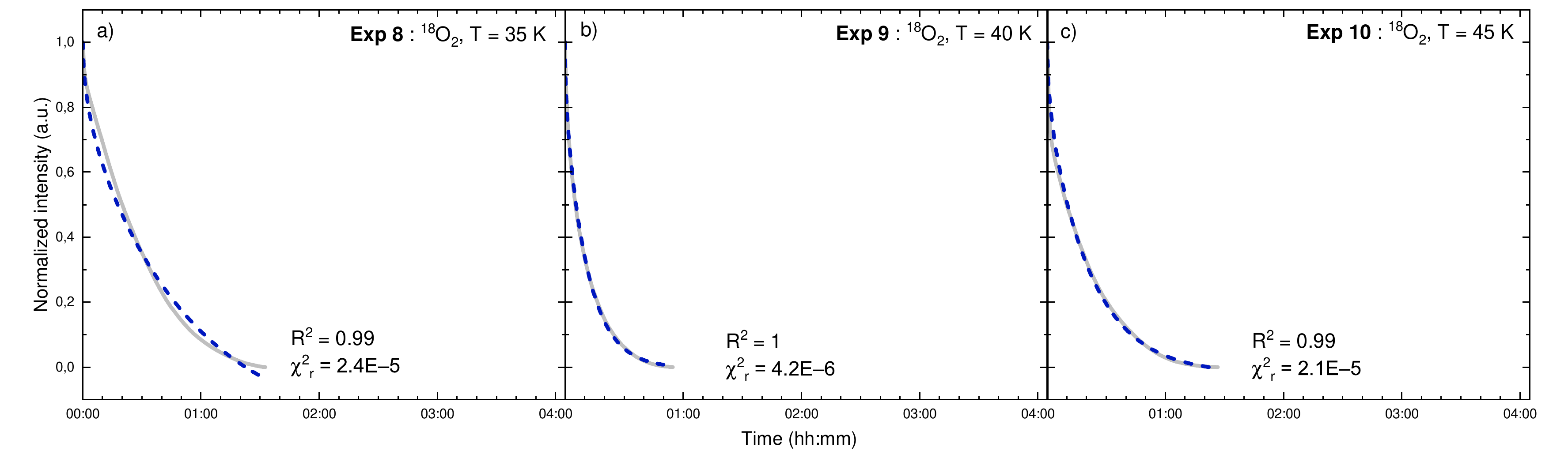}}
  \caption{Results of the fits to the Fickian model (blue) for all experimental data (grey) recorded during the four hour isothermal phase: (a)–(c) correspond to the fits of experiments 8 to 10 conducted at 35 K, 40 K and 45 K respectively, with a 40 ML water-ice matrix. Data is truncated once diffusion ceases, i.e., when the plateau is reached.}
  \label{Fig. C.2}
\end{figure*}

All isotopic experiments were also fitted to Eq.~\ref{eq6} using the fixed parameters listed in Table~\ref{table:C.1} The resulting fits are shown in Fig.~\ref{Fig. C.2}, and the extracted diffusion coefficients are reported in Table~\ref{table:C.1}. The fits show excellent agreement with the experimental data, as indicated by $R$\textsuperscript{2} values above 0.99 and $\chi$\textsuperscript{2} values below 2.4 $\times$ 10\textsuperscript{$-$5}. The extracted diffusion coefficients are within the expected order of magnitude, ranging from (0.5 $\pm$ 3.5) to (2.3 $\pm$ 4.4) $\times$ 10\textsuperscript{$-$15} cm\textsuperscript{2} s\textsuperscript{$-$1}, but exhibit large uncertainties. Consequently, no further analysis was conducted, as the mechanism cannot be fully constrained. Nevertheless, the consistent fitting of both these isotopically labeled control experiments and the major isotopologue experiments confirms that we are probing the surface diffusion of O\textsubscript{2} through ASW from the bottom to the ice surface.

\begin{table*}
\centering
\caption{Fixed parameters used for Fick’s diffusion law analysis of all initially layered \textsuperscript{18}O\textsubscript{2}–H\textsubscript{2}O experiments, along with their extracted diffusion rate.}                 
\label{table:C.1}    
\centering                        
\begin{tabular}{c c c c c}      
\hline\hline               
Experiment & $A_{0}$ (10\textsuperscript{$-$7} mbar s\textsuperscript{$-$1}) \tablefootmark{a} & $h$ (10\textsuperscript{$-$6} cm) & $D$ (10\textsuperscript{$-$15} cm\textsuperscript{2} s\textsuperscript{$-$1}) & $s$ (10\textsuperscript{$-$8} mbar s\textsuperscript{$-$1}) \tablefootmark{b} \\         
\hline                      
   8 & 0.6 & 2.0 $\pm$ 0.4 & 0.5 $\pm$ 3.5 & $-$1.2 $\pm$ 18.3 \\    
   9 & 1.0 & 2.0 $\pm$ 0.4 & 2.3 $\pm$ 4.4 & $-$0.1 $\pm$ 4.4 \\
   10 & 1.0 & 2.1 $\pm$ 0.4 & 1.3 $\pm$ 2.3 & $-$0.2 $\pm$ 4.8 \\
\hline                                  
\end{tabular}
\tablefoot{\tablefoottext{a}{All $A_{0}$ have an uncertainty of $\pm$ 1 $\times$ 10\textsuperscript{$-$10} mbar s\textsuperscript{$-$1};}
\tablefoottext{b}{The offset parameter $s$ is negligible in our study, being nearly two orders of magnitude smaller than $A$\textsubscript{0}.}
}

\end{table*}

\begin{figure}[h!]
  \resizebox{\hsize}{!}{\includegraphics{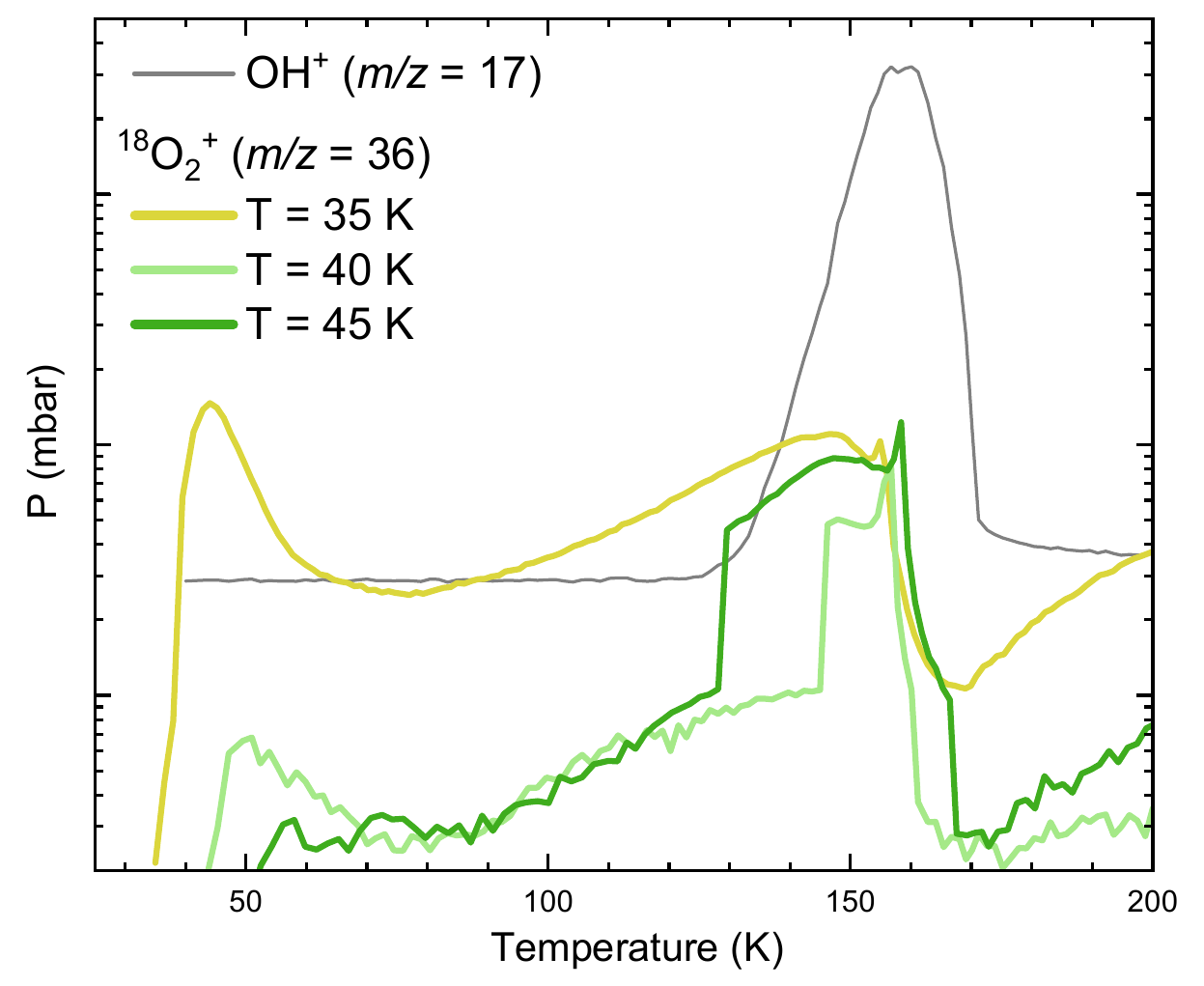}}
  \caption{QMS signals recorded during the TPD phase at $m/z$ = 17 (OH\textsuperscript{+}) and $m/z$ = 36 (\textsuperscript{18}O\textsubscript{2}\textsuperscript{+}) for experiments 8 to 10 at different isothermal temperatures (35 K to 45 K with 40 ML of H\textsubscript{2}O). Considering the signal at $m/z$ = 17, showing water desorption peaking at around 150 K, was the same for every experiment, only the water profile of experiment 9 (T = 40 K) is shown for clarity.}
  \label{Fig. C.3}
\end{figure}

Figure~\ref{Fig. C.3} shows the QMS signal recorded during the TPD of the isotopic experiments. Similar to the non-isotopic experiments, two desorption features are observed at $m/z$ = 36. While the first desorption peak exhibits a behavior comparable to that of regular O\textsubscript{2}, the absence of a clear trend in the second desorption peak suggests that contamination of the water ice may affect the amount of entrapped \textsuperscript{18}O\textsubscript{2}. Indeed, if regular O\textsubscript{2} from the air is already present in the water ice matrix and occupies the most favorable trapping sites, it reduces the number of available sites for \textsuperscript{18}O\textsubscript{2} entrapment. However, once corrected for contamination (Sect.~\ref{sec:2.3.2}), the entrapment efficiencies of \textsuperscript{18}O\textsubscript{2} in ASW (23\%, 20\%, and 19\%, at 35 K, 40 K, and 45 K, respectively) are consistent with those of regular O\textsubscript{2}, sharing the same lower limit of about 20\% of entrapment at high temperatures.

\end{appendix}
\end{document}